\documentclass[onecolumn]{aastex62}
\title{Ices Paper}

\newcommand\aastex{AAS\TeX}

\newcommand{\choh}{CH$_3$OH }
\newcommand{\hto}{H$_2$O }
\newcommand{\chisq}{$\chi^2$ }
\usepackage{xcolor}
\usepackage{url}
\usepackage{longtable} 
\usepackage{times}
\usepackage{float}
\usepackage{wrapfig}
\usepackage{rotating}
\usepackage{amssymb}
\usepackage{xcolor}
\usepackage{dejavu}
\usepackage{colortbl,booktabs,array,tabularx}
\usepackage{lipsum,graphicx,float}
\usepackage[caption=false]{subfig}
\usepackage{graphicx}
\usepackage{rotating}
\usepackage{scrextend}
\usepackage{hyperref}

\newcommand{\arcs}{\mbox{\ensuremath{.\!\!^{\prime\prime}}}}

\received{May 28, 2020}
\accepted{October 7, 2020}

\submitjournal{Astrophysical Journal}

\shorttitle{\aastex\ Ices}
\shortauthors{Chu et al.}

\begin{document}

\title{Observations of the Onset of Complex Organic Molecule Formation in Interstellar Ices}

\correspondingauthor{Laurie E Chu}
\email{lurban@ifa.hawaii.edu}

\author[0000-0002-1437-4463]{Laurie E. U. Chu}
\thanks{Visiting Astronomer at the Infrared Telescope Facility, which is operated by the University of Hawaii under contract NNH14CK55B with the National Aeronautics and Space Administration.}
\affiliation{Institute for Astronomy, 640 N. Auhoku Pl. \#209, Hilo, HI 96720 USA}

\author{Klaus Hodapp}
\thanks{Visiting Astronomer at the Infrared Telescope Facility, which is operated by the University of Hawaii under contract NNH14CK55B with the National Aeronautics and Space Administration.}
\affiliation{Institute for Astronomy, 640 N. Auhoku Pl. \#209, Hilo, HI 96720 USA}

\author{Adwin Boogert}
\affiliation{Institute for Astronomy, 2680 Woodlawn Dr., Honolulu, HI 96822 USA}

\begin{abstract}
Isolated dense molecular cores are investigated to study the onset of complex organic molecule formation in interstellar ice.  Sampling three cores with ongoing formation of low-mass stars (B59, B335, and L483) and one starless core (L694-2) we sample lines of sight to nine background stars and five young stellar objects (YSOs; A$_K$ $\sim$ 0.5 - 4.7).  Spectra of these stars from 2-5 $\mu$m with NASA's Infrared Telescope Facility (IRTF) simultaneously display signatures from the cores of H$_2$O (3.0 $\mu$m), \choh (C-H stretching mode, 3.53 $\mu$m) and CO (4.67 $\mu$m) ices.  The CO ice is traced by nine stars in which five show a long wavelength wing due to a mixture of CO with polar ice (CO$_r$), presumably CH$_3$OH.  Two of these sight lines also show independent detections of \choh.  For these we find the ratio of the CH$_3$OH:CO$_r$ is 0.55$\pm$0.06 and 0.73$\pm$0.07 from L483 and L694-2, respectively. The detections of both CO and \choh for the first time through lines of sight toward background stars observationally constrains the conversion of CO into \choh ice. Along the lines of sight most of the CO exists in the gas phase and $\leq$15\% of the CO is frozen out. However, \choh ice is abundant with respect to CO ($\sim$ 50\%) and exists mainly as a CH$_3$OH-rich CO ice layer.  Only a small fraction of the lines of sight contains \choh ice, presumably that with the highest density.  The high conversion of CO to \choh can explain the abundances of \choh ice found in later stage Class 1 low mass YSO envelopes (CH$_3$OH:CO$_r\sim$0.5-0.6).  For high mass YSOs and one Class 0 YSO this ratio varies significantly implying local variations can affect the ice formation. The large \choh ice abundance indicates that the formation of complex organic molecules is likely during the pre-stellar phase in cold environments without higher energy particle interactions (e.g. cosmic rays).    


\end{abstract}



\section{Introduction}\label{sec:intro}
Dense molecular cores are rich in interstellar ices and may well probe the origins of organic and volatile material in our Solar nebula. The environments of molecular cores are shielded from interstellar radiation where gas can readily condense onto cold dust particles forming icy mantles.  The grain surface chemistry is moderated by the H/H$_2$ ratio within the cores since hydrogen is the most abundant element in these environments.  This ratio decreases rapidly with increased density within the core \citep{Hollenbach1971}.  At relatively low densities where the ratio of H/H$_2 \geq1$ the formation of \hto and mixtures with \hto ice dominate \citep{Allamandola1999}. After an initial monolayer of ice has been deposited, there is an onset of rapid ice mantle growth that begins near the edges of the cloud when photodesorption has decreased at an ice formation threshold of A$_V$=1.6 \citep{Hollenbach2009}. Then as the dust extinction increases the  photodesorption decreases allowing CH$_4$, NH$_3$, and CO$_2$ (polar component) to develop as mixtures with H$_2$O.  At very high densities and cold temperatures where H/H$_2$ ratios are substantially less than 1, CO can accrete directly from the gas \citep{Allamandola1999}.  The subsequent phase is a nearly complete freeze out of CO at very cold temperatures and high densities (n$\geq$10$^5$ cm$^{-3}$) where the H/CO gas ratio is considerably enhanced.  This leads to the hydrogenation of CO ice on grain surfaces forming H$_2$CO and \choh on short timescales on the order of $\sim$few~$\times$~10$^4$ years \citep{Cuppen2009}.

Beyond this phase even more complex molecules can develop.  Laboratory studies demonstrate that complex organic molecules (COMs) can form from simple ices such as methanol (CH$_3$OH) after exposure to ultraviolet (UV) photons and cosmic ray particles \citep{Oberg2009}.  After heavy processing organic residues similar to insoluble organic material found in carbonaceous meteorites appears \citep{Greenberg1995} and may be of biologic importance \citep{Bernstein1995, Munoz2004}.  Alternatively \citet{Fedoseev2017} use laboratory experiments to demonstrate that complex molecule formation can occur at low temperatures in a period well before thermal and energetic processing is dominant.  Clearly, the physical conditions promoting or inhibiting ice formation (e.g., density, time scales, stellar feedback) are critical, but are presently poorly observationally constrained.

Observing field stars behind molecular cores in the infrared continuum reveals absorption bands caused by ices in the core.  Many of the absorption features display complex profiles as a result of dipole interactions in ice mixtures, which can change due to the ice structure (thermal history), and the grain shape and size.  The ices \hto (3.0 $\mu$m), CO (4.67 $\mu$m), and CO$_2$ (4.27 $\mu$m) have relatively well understood profiles.  The first detection of these three ices was in the Taurus Molecular Cloud (TMC) \citep{Whittet1983, Whittet1985, Whittet1998}.  Later on \choh (3.53 $\mu$m C-H stretching mode) was discovered by \citet{Boogert2011} and \citet{Chiar2011}. 

Most of these ices are relatively difficult to observe from the ground due to telluric features and high sky background noise in the infrared.  To detect ices in a line of sight through a molecular core we require infrared-bright giant background stars, which has limited the sample in which to probe the environments where ices form.  Both small isolated cores and larger molecular clouds have been studied for ice features and there are significant variations between each of them.  For example, the extinction threshold for \hto ice to form in the Ophiuchus (Oph) cloud is  $A_{\rm V}\sim10-15$ mag while the Pipe Nebula has a threshold of $A_{\rm V}=5.2\pm6.1$ mag which are both higher than that of the Taurus Molecular Cloud (TMC) or Lupus ($A_{\rm V}=3.2$ and 2.1 mag respectively) \citep{Tanaka1990, Whittet2001, Boogert2013, Goto2018}.  This could be due to nearby hot stars in Oph creating a higher interstellar radiation field and thus \hto ice photo-desorbs at the cloud edge \citep{Hollenbach2009}.  If a cloud hosts a Young Stellar Object (YSO) then shocks or radiation from within the core may also suppress ice mantle formation.  Thus it is desirable to study a variety of clouds at different evolutionary stages to better understand what conditions promote or restrict ice growth.

The CO freeze out and subsequent \choh formation may be particularly sensitive to environmental conditions.  The CO ice formation sets the gas phase H/CO ratio because as CO freezes out, this ratio increases, and that in turn sets the rate of hydrogenation of CO ice on dust grains to produce CH$_3$OH \citep{Cuppen2009}.  The CO to \choh conversion progresses with time but the dust temperature and gas density can impact the effectiveness of the process. According to \citet{Cuppen2009} figures 1 and 2, the \choh production after 10$^5$ years at 15.0 K and with a density of n$_H\sim$10$^4$ cm$^{-3}$ is an order of magnitude lower than at temperatures of 12.0 K.  Increasing the density to n$_H\sim$10$^5$ cm$^{-3}$, the \choh production increases and at 15.0 K it is as much as for n$_H\sim$10$^4$ cm$^{-3}$ at 12.0 K.  Therefore the environment of the star formation can greatly impact the ice freeze-out.  Large variations of the \choh abundances have been measured with N(CH$_3$OH)/N(H$_2$O)~$\leq$~3\% toward TMC background stars \citep{Chiar1995}, $\sim$10\% for isolated dense cores \citep{Boogert2011}, and upper limits between 3-8\% for Lupus \citep{Boogert2013}.
 

The absorption profile of CO may provide more information on the hydrogenation process of CO into CH$_3$OH and give insight into the efficiency of \choh formation by CO hydrogenation.  Solid CO has been found to reside in three molecular environments and as a result the ice spectral feature is made of three separate components \citep{Pontoppidan2003}.  The central feature peaking at 4.67 $\mu$m is attributed to apolar CO and similarly the blue component peaking at 4.664 $\mu$m is likely a mixture of CO with other apolar species such as CO$_2$, N$_2$, or O$_2$ \citep{Elsila1997}.  The red component is instead explained by a mixture of CO with a polar ice \citep{Sandford1988}.  It is possible that with the large amount of \hto ice the CO ice mixes at the H$_2$O-CO interface where the CO can migrate into the \hto pores and cause the polar component.  However if this were the case we would observe the 4.647 $\mu$m band from dangling OH bonds \citep{Sandford1988}.  Instead the polar wing is most likely due to \choh since it is the second most abundant polar molecule in interstellar ices \citep{Cuppen2011, Penteado2015}.  This mixture, along with an independent, direct detection of \choh can give us more insight into the \choh formation conditions and efficiency.

In the following section (Section \ref{sec:targets}), we present the targets in our sample and provide background on the molecular clouds to put our ice measurements into context.  In Section \ref{sec:obs}, the observational and reduction methods are described, and Section \ref{sec:results} explains the modeling of the spectral absorption ice features and demonstrates the relation of the optical depth for ices with the extinction.  In Section \ref{sec:Discussion}, we analyze the relationship between the CO and \choh ices and how we can learn about the onset of complex organic molecules.

\section{Target Selection and Molecular Core Context}\label{sec:targets}

We have chosen to study the ice abundances in four nearby ($\leq$ 250 pc) small ($\sim$0.2-1 pc) dense molecular cores.  These cores were selected because they represent different stages of evolution where one is collapsing (L694-2), two have class 0 Young Stellar Objects (YSOs; B335 and L483), and one is quiescent with several later stage YSOs (B59).  The first three are isolated cores and B59 is part of the larger Pipe Nebula. The different cores enable us to study how the environment can impact the ice chemistry. Because they are relatively nearby, the lines of sight through the cores can be sampled by observing the background stars without contamination from foreground stars.  Each is positioned against the galactic bulge where there is a high density of stars.  This increases the likelihood to find bright enough background stars at different extinction levels to sample the molecular cores.  The background field of galactic bulge stars is also beneficial because we expect that they are mostly late type stars making it easier for spectral classification.

\begin{figure*}[!htb]
\centering
\includegraphics[width=0.75\textwidth]{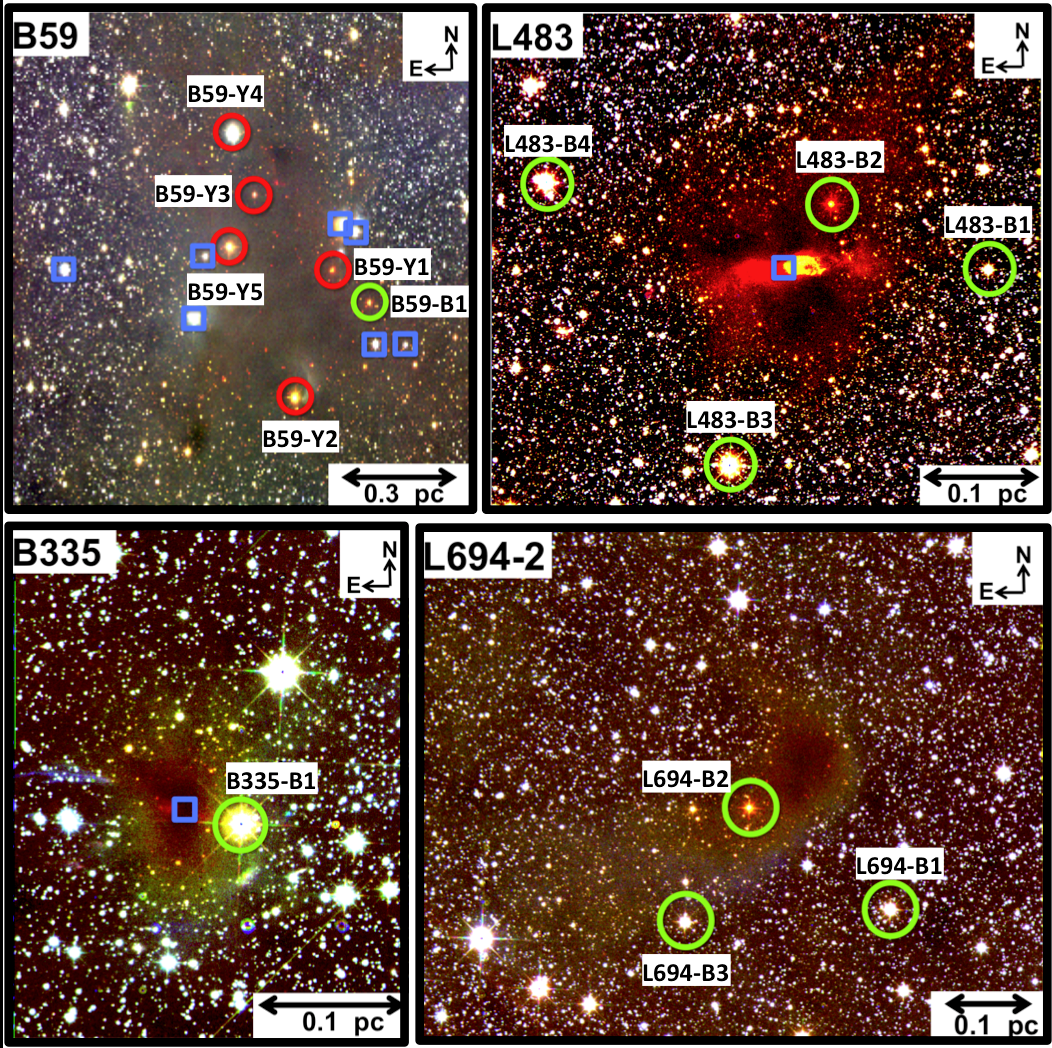}
\caption{Color images for the molecular cores in which we study ice absorption features.  These were made by combining JHK colors from the UKIRT  WFCAM instrument. The following are the total exposure times used for each cloud: \textit{B59}: J=1.0 hr, H=1.2 hr, K=1.4 hr, \textit{L483}: J=5.6 hr, H=2.8 hr, K=2.8 hr, \textit{B335}: J=4.0 hr, H=2.8 hr, K=3.2 hr. \textit{L694-2}: J=3.2 hr, H=2.0 hr, K=2.0 hr.  Stars circled in green represent background targets and red circles are YSO candidates in our sample.  Blue squares show where other nearby YSOs are located in relation to targets in our sample.  Size scales are shown for the quoted distances in the text for each core. }
\label{fig:color images}
\end{figure*}

We have identified 14 stars to sample the four molecular cores.  In Table \ref{tab:sources}, each target is given an alias that will be used throughout this paper including the name of the core, and distinguishing whether the target is a background star or YSO.  They cover a wide range of extinction values (5 $\leq$ A$_V$ $\leq$ 50) and are brighter than 10.5 magnitude in the WISE Channel 1 band (3.35 $\mu$m) and brighter than 10 magnitudes in the WISE Channel 2 band (4.6 $\mu$m).  Any stars fainter than this limit would not produce a high enough signal to noise because of the high sky background noise beyond 4 $\mu$m where some of the ice absorption features are. Two of the fourteen stars (B59-B1 and L483-B2) have been previously observed by \citet{Boogert2011} but did not include coverage from $\sim$4--5 $\mu$m missing the CO ice feature (4.67 $\mu$m). Five of the stars are YSOs in Barnard 59 (Section \ref{sec:B59}) and were included in our sample because they displayed high levels of extinction (A$_V>20$), or in the case of B59-Y4 because it samples a line of sight further from the core, providing a good spatial coverage.  We did not choose stars that sampled very similar lines of sight through the core.  YSOs in B335 and L483 are heavily reddened in the center of the core and thus too weak for L and M band spectroscopy with SpeX/IRTF. The observed stars are summarized in Table \ref{tab:sources}. To identify the location of our targets relative to the overall molecular core structure and relation to nearby YSOs, Figure \ref{fig:color images} shows composite JHK color images created from the United Kingdom Infrared Telescope (UKIRT) Wide Field Camera (WFCAM).  The images combined stacked images with total exposure times listed in the caption.  They were reduced using the standard WFCAM pipeline, and extinction maps will be derived in future work (Chu et al. in prep).  The reddest regions of the images show the densest parts of the core, and the stars trace extinctions just outside the densest regions of the core. At any higher extinctions, the background stars are too faint for sufficient data quality.  Below we provide details on each core for context on the selected background stars and describe how the extinction values are calculated in Table \ref{tab:sources}.   

\begin{table*}[h!t]
\tabcolsep=0.1cm
\tabletypesize{\small}
\centering
\caption{Background Star Sample}
\label{tab:sources}

\begin{tabular}{cccrrrrrrr}
\tableline\tableline
Source 2MASS J	& Cloud & Alias\footnote{Name given to identify the cloud the star samples, and whether it is a Background target (B) or YSO (Y)} &	WISE Ch 1			&	WISE Ch 2			&	Date Obs	&	Int Time\footnote{On sky without overheads included}	&	A$_V$	\\
\tableline
17111501-2726180	& B59	& B59-B1	&	9.55$\pm$0.02	&	8.92$\pm$0.02	&	5/24/17	&	56 min	&	36$\pm$1.8 	\\
17111827-2725491	& B59 & B59-Y1	&	9.80$\pm$0.02	&	8.14$\pm$0.02	&	5/21/16	&	40 min	&	49.8$\pm$2.2	\\
17112153-2727417	& B59 & B59-Y2	&	7.85$\pm$0.02	&	6.50$\pm$0.02	&	5/23/17	&	45 min	&	22.6$\pm$0.6	\\
17112508-2724425	& B59 & B59-Y3	&	10.40$\pm$0.02	&	9.11$\pm$0.02	&	5/22/16	&	80 min	&	30.3$\pm$0.7	\\
17112701-2723485	& B59 & B59-Y4	&	8.46$\pm$0.02	&	7.57$\pm$0.02	&	5/23/17	&	48 min	&	14.3$\pm$0.7	\\
17112729-2725283	& B59 & B59-Y5	&	8.10$\pm$0.22	&	7.12$\pm$0.14	&	5/24/17	&	40 min	&	22.6$\pm$0.6	\\
18171765-0439379	& L483 & L483-B1	&	9.61$\pm$0.02	&	9.65$\pm$0.02	&	8/16/16	&	60 min	&	5.7$\pm$1.0	\\
18172690-0438406	& L483 & L483-B2	&	9.25$\pm$0.02	&	8.26$\pm$0.02	&	5/26/17	&	60 min	&	40.4$\pm$2.1	\\
18173285-0442271	& L483 & L483-B3	&	6.95$\pm$0.06	&	6.66$\pm$0.02	&	8/15/16	&	40 min	&	8.2$\pm$1.0	\\
18174365-0438205	& L483 & L483-B4	&	7.03$\pm$0.06	&	6.54$\pm$0.03	&	8/15/16	&	40 min	&	7.8$\pm$1.1	\\
19365867+0733595	& B335 & B335-B1	&	7.11$\pm$0.03	&	7.00$\pm$0.02	&	5/21/16	&	40 min	&	15.5$\pm$1.1	\\
19405855+1054527	& L694-2 & L694-B1	&	8.37$\pm$0.02	&	8.34$\pm$0.02	&	5/26/17	&	58 min	&	7.8$\pm$1.1	\\
19410754+1056277	& L694-2 & L694-B2	&	8.87$\pm$0.02	&	8.27$\pm$0.02	&	5/22/16	&	52 min	&	27.2$\pm$1.1	\\
19411163+1054416	& L694-2 & L694-B3	&	8.83$\pm$0.02	&	8.79$\pm$0.02	&	8/14/16	&	40 min	&	7.0$\pm$1.1	\\

\tableline
\end{tabular}
\end{table*}

\subsection{Barnard 59} \label{sec:B59}

Barnard 59 (hereafter B59) is part of the Pipe Nebula at a distance of 163 pc \citep{Dzib2018} with a total mass of $\sim$30 M$_\odot$ \citep{DuarteCabral2012}.  Column density maps of the Pipe Nebula from \citet{Peretto2012} using the Herschel PACS/SPIRE instrument show that the B59 region has the highest extinction levels within the Pipe Nebula (above A$_V\simeq8$), which corresponds to the extinction threshold where protostar formation is believed to take place (e.g. \citet{Heiderman2010}, \citet{Andre2010}).  Indeed, B59 is the only active star forming region known in the Pipe Nebula \citep{Onishi1999, Forbrich2009} with a small cluster of YSOs near the center, of which some are observed in our sample.  It is suggested that the outflows from the small number of protostars assist in sustaining supersonic turbulence within the cloud on sub-parsec size scales preventing it from collapsing and thus it is considered a stable core \citep{DuarteCabral2012}.  This stability of the core and the presence of YSOs make this core an interesting stage of evolution in which to study ice formation.  Six of our targets sample B59, one is a background star while the remaining are YSOs.

\subsection{LDN 483}\label{sec:L483}

LDN 483 (hereafter L483) is a dark isolated cloud about 200-250 pc away \citep{Dame1985, Goodman1993, Felli1992, Jorgensen2002} and harbors a variable protostar, IRAS 18148-0440 \citep{Parker1988}.  The star is deeply embedded with A$_V>$~70 mag \citep{Fuller1995}. Studies classify the protostar as transitioning from a Class 0 YSO to a Class I YSO because the SED is characteristic of a Class 0, while the outflow and NIR reflection nebula implies that it is a Class I YSO \citep{Tafalla2000}. It is also noteworthy that significant enhancements of \choh and SiO are absent toward the protostar implying the outflows are more evolved than in Class 0 protostars \citep{Tafalla2000}. \citet{Connelley2009} find the surrounding envelope is cool which is also indicative of a Class I object that is still deeply embedded.  The star is accompanied by prominent H$_2$ jets that are typical in this stage of stellar formation \citep{Fuller1995}. The mass of the protostar is between 0.1-0.2M$_{\odot}$ \citep{Oya2017} with a bolometric luminosity of 10-14 L$_{\odot}$ \citep{Ladd1991,Tafalla2000}.

This cloud is particularly interesting because over time scales on the order of months, morphological changes have been observed in the cloud where features such as clumps and knots in the nebula significantly change in brightness.  One knot even disappears and then reappears in the same location \citep{Connelley2009}. These changes are seen in the near-IR within the bipolar outflows oriented in the east and west directions.   It is believed the variability is due to a change in the illumination of the nebula caused by opaque clouds potentially part of a circumstellar disk that are within $\sim$1 AU of the protostar.  On longer timescales, the reflection nebula is decreasing in brightness and was observed to decrease by two magnitudes in the course of a year \citep{Connelley2009}.  This timescale is shorter than the amount of time for light to cross the nebula, proving that the changes in the illumination of the nebula are quick.    We include four background targets to trace this core.

\subsection{Barnard 335}

The Bok globule, Barnard 335 (hereafter B335) is an isolated optically opaque cloud located about 150 pc away \citep{Stutz2008}.  This core appears to be less evolved than L483 with a Class 0 YSO deeply embedded in the cloud \citep{Frerking1982} and has been detected in the far-infrared \citep{Keene1983} and is bright in the submillimeter regime \citep{Chandler1990}.  The mass and luminosity of the protostar are lower than the embedded YSO in L483 \citep[0.02-0.06 M$_\odot$ and 0.72 L$_\odot$,][]{Imai2019,Evans2015}.  Both L483 and B335 have infalling mass \citep{Zhou1993, Chandler1993} but B335 only displays low levels of rotation while in L483 there are clear rotational features \citep{Jacobsen2019}.  Neither however show signs of a Keplerian disk down to 10-15 AU \citep{Jacobsen2019,Bjerkeli2019}.

Based on CO maps, a well collimated outflow is evident lying close to the plane of the sky \citep{Chandler1993}.  \citet{Hodapp1998} found that the outflows are more representative of an older protostar because the outflow has broken through the globule on both sides and there are infrared emission features near the heavily embedded central source.  Further observations at 8 $\micron$ by \citet{Stutz2008} revealed a shadow extending $\sim$3000-7500 AU with A$_V>$100.  From $^{12}$CO measurements, a rotating structure on scales of $\sim$ 10,000 AU that is coaxial with the central circumstellar disk is present.  It is believed that this rotating structure and shadow are related.  This is one of the most well studied Class 0 YSOs and proves to be a good example of a collapsing core, and in our sample, we have one background target tracing this cloud.

\subsection{LDN 694-2}

The cloud LDN 694-2 (hereafter L694-2) is an isolated mostly round dense core with a lower extinction extended "cometary tail-like" feature.  It is situated approximately 250 pc away \citep{Tomita1979, Kawamura2001} and is the only starless core we sample.  Several studies have worked to see if there is a protostar in this cloud \citep{Harvey2002, Harvey2003, Evans2003}, but no such source was found.  If there is such a source that is invisible still, the luminosity is less than about 0.3$L_\odot$ and no compact disk was found \cite{Harvey2003}. Mapping the N$_2$H$^+$(1-0) emission, \citet{Lee2001} discovered a compact centrally condensed core, and \citet{Williams2006} estimate the mass of the core to be $\sim 1M_\odot$ and predict that a point mass will form in a few 10$^4$ yr.  It has been determined that the core is collapsing with a strong infall with infall speeds increasing toward the center \citep{Lee2004}.  This provides an excellent target to understand the ice mantle chemistry at a time before a YSO is present, and thus we sample three lines of sight through the core.

\subsection{Determining Extinction Values of Background Stars}

Using the 2MASS JHK (or HK if there is no detection in J band) we employ the NICER technique \citep{lombardi2001} to calculate the extinction for each star.  This technique allows more than two photometric bands to determine the extinction and is calibrated using a control field without extinction.  Our control field is simulated for each cloud using the TRILEGAL algorithm for simulating stellar photometry for any field in the galaxy \citep{Girardi2005}.  Using this field we develop a covariance matrix and average stellar color for each core and use a reddening vector from \citet{Indebetouw2005}.  The colors and corresponding spectral types for each core are in Table \ref{tab:int-color}, and the extinctions for each background star are reported in Table \ref{tab:sources} (A$_V$).  This method will be used more generally for extinction mapping purposes in Chu et al. (in prep), where further details of the procedure will be outlined. 

\begin{table*}[h!t]
\tabletypesize{\small}
\centering
\caption{Average Intrinsic Color of Background Stars}
\label{tab:int-color}
\begin{tabular}{lccc}

\tableline\tableline

Cloud & J-H & H-K & Spectral Type \footnote{Corresponding to subclass I-III taken from the H-K color \citep{Koornneef1983}}\\
\tableline
B59 & 0.440 & 0.109 & G3-K0 \\
L483 & 0.805 & 0.299 & M4-M5 \\
B335 & 0.530 & 0.129 & K0-K2 \\
L694-2 & 0.613 & 0.174 & K3-K5 \\

\tableline

\end{tabular}
\end{table*}

\section{Observations and Data Reduction}\label{sec:obs}

We have obtained stellar spectra from the NASA Infrared Telescope Facility (IRTF) SpeX instrument from $\sim$2 $\mu$m to $\sim$5 $\mu$m for all 14 stars in Table \ref{tab:sources}. This spectral range allows for measurements of the H$_2$O (3.0 $\mu$m), CO (4.67 $\mu$m), CH$_3$OH (3.53 $\mu$m C-H stretching mode), and OCN$^-$ (4.62 $\mu$m) ice features.

\subsection{Observational Setup and Reduction}

 Using the SpeX instrument on NASA's IRTF, we observed with the Long Cross Dispersed long-wavelength mode (LXD{\_}long) to cover a wavelength range of 1.98 - 5.3 $\mu$m with R $\sim$800-950 and a slit size of 0$\arcs$8 \citep{Rayner2003}.  Observations were taken over three consecutive semesters on eight different nights with total integration times ranging from 40 minutes up to 80 minutes for the different stars as listed in Table \ref{tab:sources}.  Since detecting the 4.67 $\mu$m CO feature requires longer integration than the H$_2$O feature, the integration time was based on getting a Signal to Noise of 100 around the CO feature.  This is high enough to detect an optical depth, $\tau\sim$0.03 at a 3$\sigma$ detection limit.  Because of the high background noise at these wavelengths, short exposure times of 10 seconds were necessary with multiple coadds.  The observations used an ABBA nod pattern, and the slit was aligned to the parallactic angle. Nearby A0V standard stars at similar airmass to the observed target were also observed along with the IRTF standardized calibrations consisting of arcs for wavelength calibration and flats.  The arcs used a thorium-argon lamp for the wavelength calibration spectra while the flats used an internal quartz lamp.  Calibrations were taken either at the start or end of our observation time.  

The spectra were reduced using the IRTF SpeX standard reduction pipeline: Spextool \citep{Cushing2004}.  We first provided the input of the calibration frames and determined the extraction apertures.  After background subtraction and flat-fielding, we extracted the data and merged the A and B beam observations to have individual calibrated frames.  Then using the tool \texttt{Xcombspec}, the observations were merged.  In cases where the signal to noise was too low on an individual frame, the tool could not extract the spectra individually and so the spectral images were first combined before extraction.  We divided over the A0V standard, and using \texttt{Xtellcor}, the telluric features were corrected. Subsequently all of the orders were combined.  The final spectrum was trimmed where there were strong telluric features particularly between 2.56 to 2.84 $\mu$m and 4.18 to 4.50 $\mu$m.  


\section{Modeling Spectra and Results}\label{sec:results}

To separate the interstellar and photospheric spectral features, we model the continuum and ice absorption features for each spectrum (Figures \ref{fig:H2O_ice} \& \ref{fig:H2O_ice_YSO}). Subsequently, we put the spectra on an optical depth scale and then calculate the peak optical depth for H$_2$O, CO, and CH$_3$OH ice.  We use methods outlined in \citet{Boogert2011, Boogert2013}, and the details of those methods are provided below.

\subsection{Modeling the Stellar Spectrum}\label{sec:modeling}

Modeling the continua of the background stars requires both spectroscopic and broad-band photometric data.  We first calibrate the IRTF spectra by convolving them with the 2MASS K-band filter profile and then multiplying them to match the K band photometry from the 2MASS database \citep{Skrutskie2006}.  The spectra do not cover shorter wavelengths, and the errors in K band are typically small.  In order to fully model the targets, we also obtained the J and H photometric bands from 2MASS and the WISE Channels 1 and 2 photometry \citep{Wright2010} for each star.  

Using both the spectral and photometric points, we model the stars with the full IRTF database of observed stellar spectra \citep{Rayner2009}.  Our models consider three main parameters: the spectral type of the observed star, the extinction toward the star, and the \hto absorption feature.  Below it is outlined how each of these are simultaneously determined and how any dependencies are taken into account into the fitted parameters and their uncertainties:

\begin{enumerate}
\item \textit{Spectral Type}.   The CO overtone lines between 2.25-2.60 $\mu$m provide a sensitive tracer of spectral type.   The near-infrared JHK photometry and absorption features in the 3.8-4.1 $\mu$m spectral range also prove to be important for deriving a spectral type. The JHK photometry depends on the extinction as well, which is discussed in point 2 below.  A reduced \chisq fit is derived for these three values and then the three reduced \chisq values are averaged to derive an overall goodness of fit for the spectral type.


\item \textit{Continuum Extinction}.  We adopt the commonly used extinction curve from \citet{Indebetouw2005}.  Due to the steepness of the extinction curve in the 1.0-2.5 $\mu$m region, the extinction A$_K$ and its uncertainty are primarily determined by the JHK photometry and the shape of the un-reddened spectral template spectrum.  The dependency on flux values at longer wavelengths is very weak.  The 3$\sigma$ uncertainty in A$_K$ is represented by models that have a reduced $\chi^2 \lesssim5$, considering that the number of free parameters is 2 (the J, H, and K photometery, minus one fitted parameter, A$_K$; see, e.g. Table C-4 in \citet{Bevington2003}).  The reported error bars are further enhanced by including the spread across a range of best-fitting templates (see below). 


\item \textit{H$_2$O Absorption Feature}.  The \hto ice feature is a prominent broad feature at 3.0 $\mu$m.  In Section \ref{sec:water ice}, we detail the modeling of this feature where we calculate the optical depth, $\tau_{3.0}$.  The optical depth at 3 $\mu$m and its error are determined by the flux values and the observational noise in the 2.9-3.2 $\mu$m range, including dependencies on the accuracy of the baseline surrounding the ice feature.  The latter in turn depends on A$_K$ and the spectral template.  This dependency is weakened by the fact that the model spectrum is normalized to the observed data at 2.5 $\mu$m.  Considering the large number of free parameters, the 3$\sigma$ uncertainty in $\tau_{3.0}$ is represented by models that have a reduced $\chi^2 \lesssim2$.  When converting the optical depth to a column density (N(H$_2$O)), the uncertainty also incorporates the error in the band strength for the 3.0 $\mu$m feature of 10\% \citep[e.g.][]{Gerakines1995}.


\end{enumerate}

Besides the reduced \chisq values derived for the individual wavelength regions discussed above, a total reduced \chisq is calculated across the full wavelength range of 1-4 $\mu$m.  Where this total reduced \chisq is lowest, the model template is chosen as the best fit to the star.  Those templates and spectral types are listed in Table \ref{tab:model_fits} as well as the total reduced $\chi^2$.  The final errors for A$_K$ and $\tau_{3.0}$ are increased by including all of the model templates with a total reduced \chisq (from the 1-4 $\mu$m range) that is within a factor of 2 of the best template.  This represents a confidence level of at least 3$\sigma$, as mentioned above.  The final A$_K$ and $\tau_{3.0}$ with corresponding 3$\sigma$ errors are reported in Table \ref{tab:model_fits}.  In some cases the uncertainties on A$_K$ and $\tau_{3.0}$ are much smaller than expected from the total reduced \chisq values because they are not sensitive to the full 1-4 $\mu$m wavelength range and the individual parameters are not impacted by an overall poor fit.

Some of the stars  (L483-B3, L483-B4, and B335-B1) have high total reduced $\chi^2$ values and are best fit with late M giant templates.  This higher reduced $\chi^2$ is due to a poor fit beyond 3.7 $\mu$m which reveals the limitations of our models in that we do not have a good representation potentially for these late type stars.  In \citet{Boogert2011}, they similarly flagged later types and stated that the fitting parameters are more uncertain.

\subsubsection{YSO Candidate Models}\label{sec:YSO models}

Five of the targets in B59 are identified as YSO candidates in \citet{Brooke2007} (Sources \#8, \#9, \#12, \#13, and \#14 in their catalog).  Four of the five stars are identified as being in the Class II stage based on their spectral energy distributions (SEDs).  The fifth has a flat spectrum representing a transition stage between Class I and II.  There are no known disk orientations for these targets but based on models in \citet{Robitaille2006} it can be assumed that the disks in this evolutionary stage are not edge-on since there is not a strong silicate absorption feature near 10 $\mu$m.  The more evolved state of these stars combined with the disk orientation imply that any ices detected are most likely from the foreground cloud and not the protostellar envelope.

We are unable to determine suitable spectra templates from the IRTF spectral library since the templates do not include YSOs. This prevents an estimation of the extinction toward the star but we can fit the baseline with a third order polynomial in order to measure ice features.  Our polynomial fit is between 2 $\mu$m to 4.1 $\mu$m excluding the region between 2.5 to 3.7 $\mu$m due to telluric features and the H$_2$O ice feature.  The fits are used to calculate the optical depth; both the spectra and optical depths are shown in Figure \ref{fig:H2O_ice_YSO}.  

In these figures, the 2.2-2.5 $\mu$m spectral range is highlighted because for background stars strong CO absorption features are present, but that is not the case for the YSOs. Instead CO emission features in this region can be a strong indicator of a YSO from the rotation-vibration overtones tracing neutral material at temperatures above 1000 K within a few stellar radii of the young star \citep{Scoville1983, Geballe1987, Carr1989}.  We find from our YSO candidates that one star shows CO emission features in this region (B59-Y3), two are relatively featureless (B59-Y1 and B59-Y5) and the remaining two have very shallow CO absorption features.  Figure \ref{fig:H2O_ice_YSO} has the CO lines marked.  All of these stars are infrared-bright, and since they do not show strong CO absorption features, they are not late-type background stars.  They could be early-type background stars but because B59 has a background field toward the galactic bulge which is mostly populated with late-type stars then it is less likely we would have an early-type background star.  These spectra combined with the SEDs in \citet{Brooke2007} make a compelling case to confirm these stars as YSOs.

\begin{table*}[h!t]
\tabcolsep=0.1cm
\tabletypesize{\small}
\centering
\caption{Continuum Fit Parameters}
\label{tab:model_fits}
\begin{tabular}{lccccc}

\tableline\tableline															
Source 	&	\multicolumn{2}{c}{IRTF\footnote{Best fitting 1-4$\mu$m spectrum from the IRTF database of \citet{Rayner2009}}}	&	A$_K$\footnote{Errors shown are 3$\sigma$}	&	$\tau_{3.0}$\footnote{Errors shown are 3$\sigma$}	&	$\chi_{\nu}^2$(total)\footnote{Averaged Reduced \chisq as described in Section \ref{sec:modeling}}	\\
ID	&	Spectral Type	&	Template	&	(Mag)	&		&		\\
\tableline											
B59-B1 	&	M6.5III	&	HD142143	&	3.59$\pm$0.25	&	1.52$\pm$0.39	&	0.66	\\
B59-Y1 	&	YSO	&	-	&	-	&	-	&	-	\\
B59-Y2 	&	YSO	&	-	&	-	&	-	&	-	\\
B59-Y3 	&	YSO	&	-	&	- 	&	-	&	-	\\
B59-Y4	&	YSO	&	-	&	-	&	-	&	-	\\
B59-Y5 	&	YSO	&	-	&	-	&	-	&	-	\\
L483-B1 	&	K4I	&	HD201065	&	0.66$\pm$0.13	&	0.11$\pm$0.06	&	0.46	\\
L483-B2 	&	M6III	&	HD18191	&	4.68$\pm$0.28	&	3.30$\pm$0.64	&	4.75	\\
L483-B3	&	M7III	&	HD108849	&	0.79$\pm$0.09	&	0.10$\pm$0.03	&	15.10	\\
L483-B4	&	M7III	&	HD207076	&	0.88$\pm$0.11	&	0.12$\pm$0.04	&	19.44	\\
B335-B1 	&	M5.5III	&	HD94705	&	1.29$\pm$0.07	&	0.48$\pm$0.12	&	20.59	\\
L694-B1 	&	M5.5III	&	HD94705	&	0.52$\pm$0.07	&	   $<$0.09	&	4.29	\\
L694-B2  	&	M2III	&	HD120052	&	2.91$\pm$0.08	&	1.30$\pm$0.06	&	2.31	\\
L694-B3  	&	K3.5III	&	HD35620	&	0.74$\pm$0.02	&	0.28$\pm$0.05	&	0.44	\\
\tableline

\end{tabular}
\end{table*}

\subsection{Ice Feature Models}
\subsubsection{\hto Ice}\label{sec:water ice}
Determining the spectral type of the star and correcting for both the continuum shape and photospheric absorption allows us to measure the peak optical depth for the \hto ice feature.  Laboratory ice spectra is used to model the H$_2$O ice feature as was done in \citet{Boogert2013}.  Spheres with radii of 0.4 $\mu$m and optical constants of amorphous solid H$_2$O at $T=10 K$ \citep{Hudgins1993} are used to derive the absorption spectrum \citep{Bohren1983}.  This allows us to fit both the H$_2$O band profiles and depths.  The pure H$_2$O ice model does not fit the red part of the absorption profile from $\sim$3.2-3.7 $\mu$m because of a possible mixture with NH$_3$ ice and scattering on larger grains.  Our model does however fit the H$_2$O ice feature between 3.0-3.2 $\mu$m which is enough to determine the optical depth. Using the band strength of 2.0~$\times$~10$^{-16}$~cm~molec$^{-1}$,  the \hto column density can be calculated \citep{Hagen1981} with errors of $\sim$10\% as mentioned in Section \ref{sec:modeling}. Fits for each star are shown in Figure \ref{fig:H2O_ice}.  Eight of the nine background stars have H$_2$O detections above 3$\sigma$, and the 3$\sigma$ error is given as an upper limit for the remaining source, L694-B1. 

As discussed in Section \ref{sec:YSO models}, the baseline for the optical depth for YSOs are determined by a polynomial fit. We then model the optical depth of the \hto ice feature with a Gaussian function by simultaneously fitting the peak wavelength and depth using the same FWHM from the laboratory ice spectra. All five YSOs display \hto ice, the column densities are reported in Table \ref{tab:col density}.  

\begin{figure*}[!htb]
\centering
\includegraphics[scale=0.72]{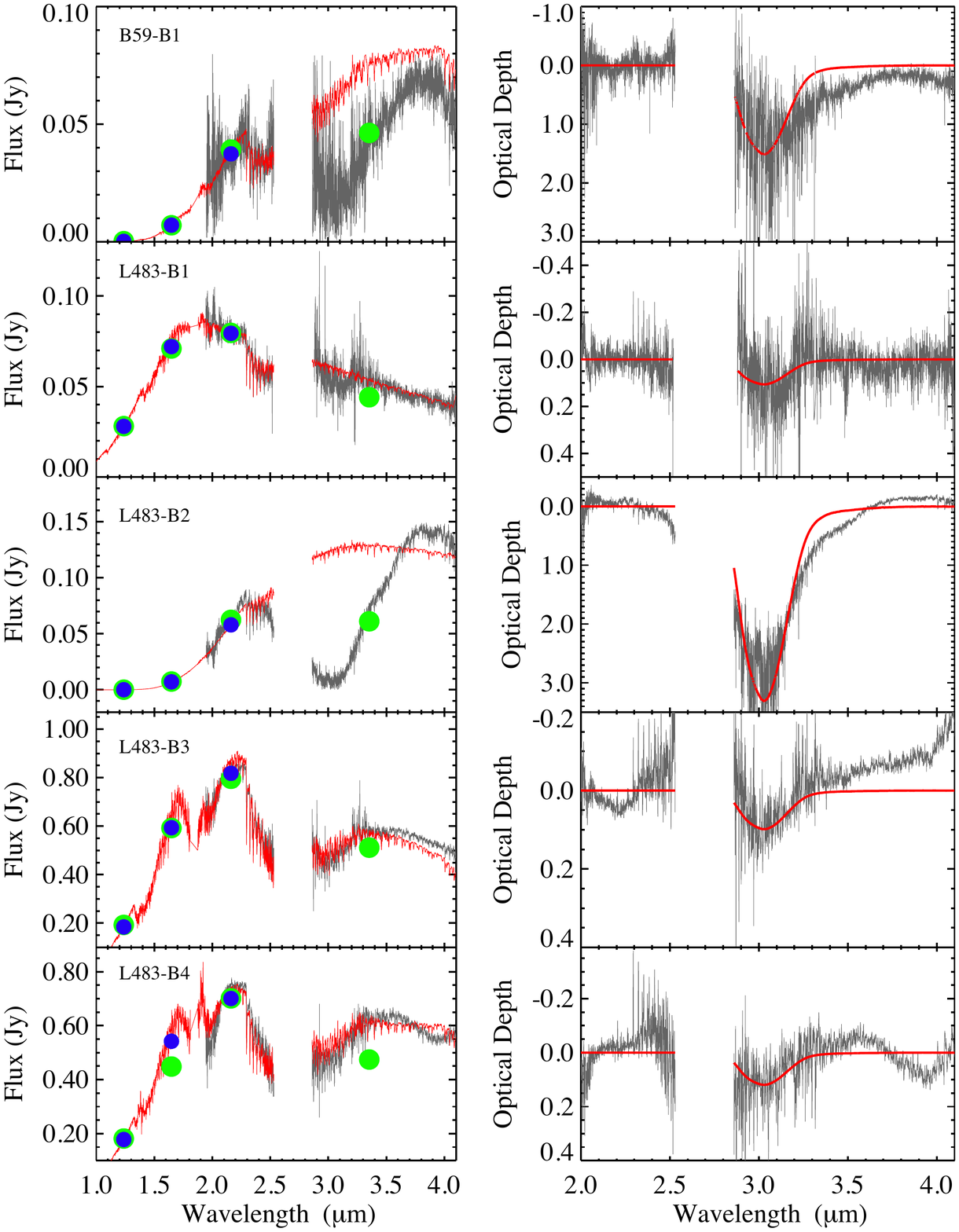}
\caption{H$_2$O ice absorption features in the K and L-band spectra for background stars.  Left panels show the flux of the star labeled in the window in grey and the red represents the best fit model star spectrum reported in Table \ref{tab:model_fits}, including the effects of continuum extinction but excluding absorption by \hto ice (although both were fitted simultaneously, \hto is omitted from the left panels to highlight the significance of the 3 $\mu$m absorption feature).  Green circles show the photometry from 2MASS and WISE, while blue circles show the modeled photometry, errors are typically smaller than the circles. Modeled L-band photometry is not shown because it does not account for the wing in the H$_2$O ice feature.}  The right panels show the optical depth for the spectrum of the background star with the red line representing the modeled fit for the H$_2$O features using laboratory measurements with the derived $\tau_{3.0}$ reported in Table \ref{tab:model_fits}. 
\end{figure*}

\begin{figure*}[!htb]
\ContinuedFloat\centering
\includegraphics[scale=0.8]{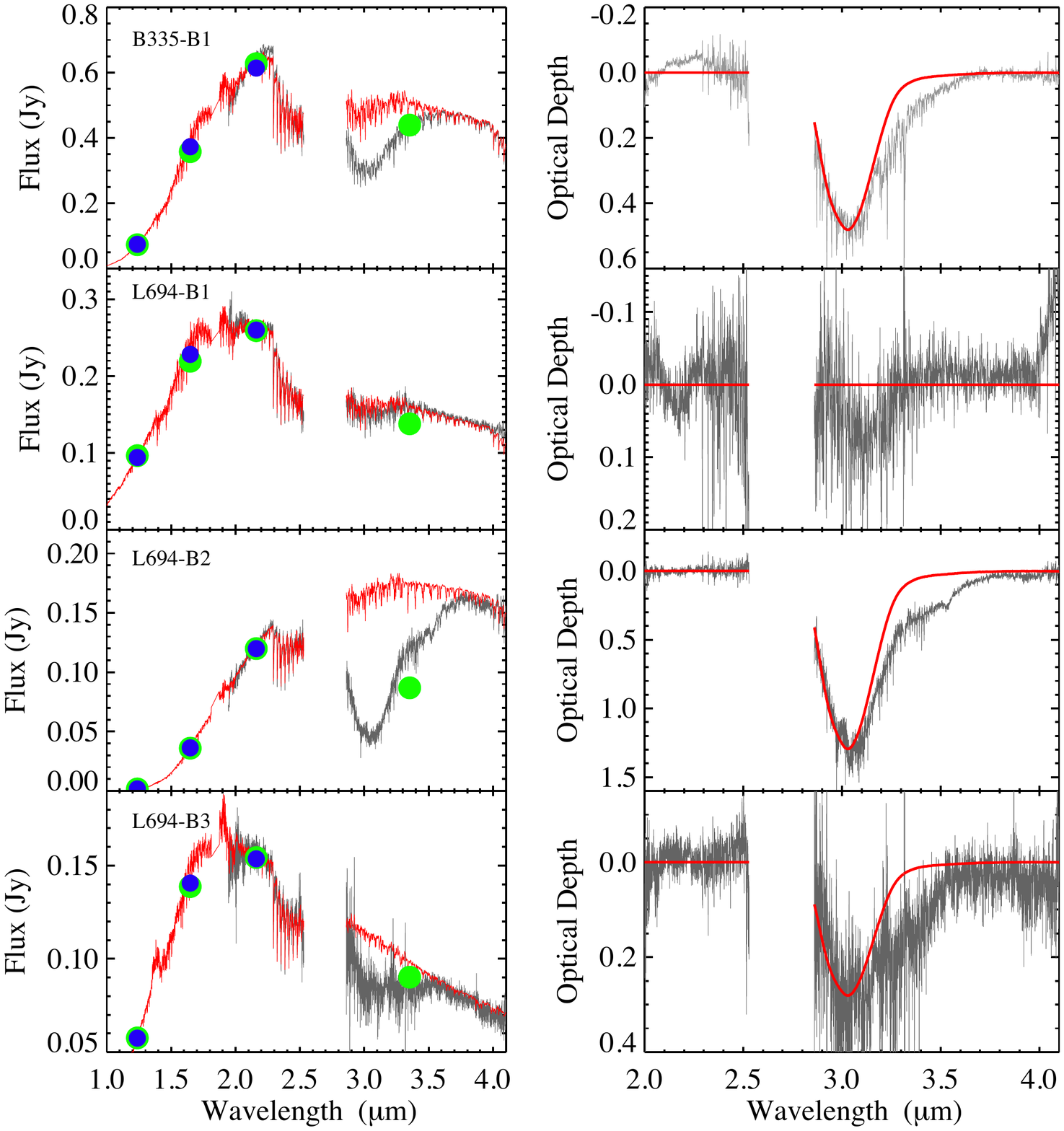}
\caption{(Continued)}
\label{fig:H2O_ice}
\end{figure*}

\begin{figure*}[!htb]
\centering
\includegraphics[scale=0.75]{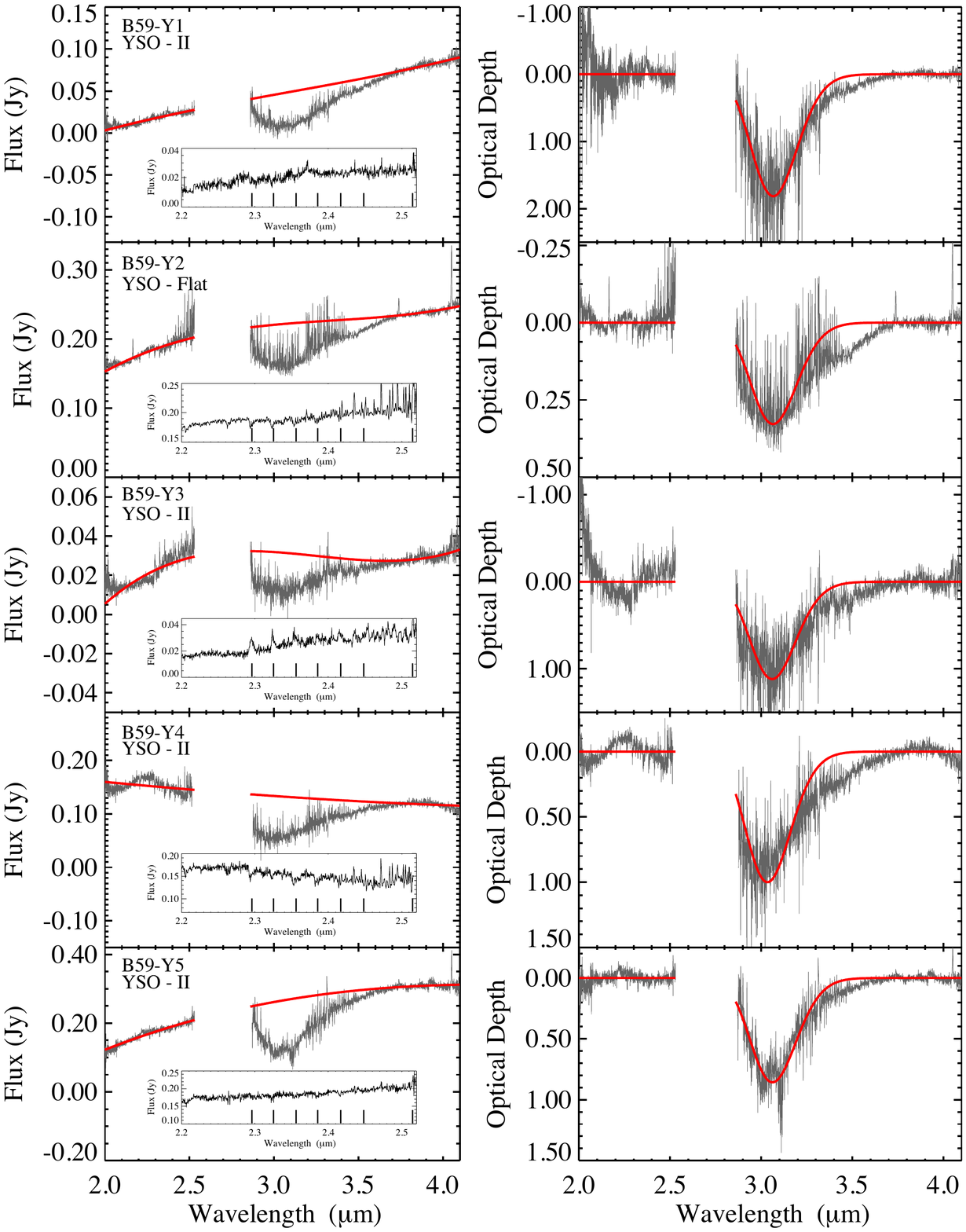}
\caption{YSO candidate spectra (grey) are shown from 2 to 4.1 $\mu$m to display the \hto ice feature.  In the left panel the spectra are fit with a polynomial function from 2-2.5 and 3.7-4.1 $\mu$m and shown in red.  Inset in each panel highlights the 2-2.5 $\mu$m region to demonstrate that some have weak CO emission and absorption features while others are rather featureless, both indicative that the star is not a background late giant star.  The CO vibration rotation bands are marked with lines representing the transitions: 2-0, 3-1, 4-2, 5-3, 6-4, 7-5, 8-6, and 9-7. The IR spectral class for each YSO is labeled \citep{Brooke2007}.  The right panels show the optical depth spectrum for each YSO with a model for the \hto feature using a Gaussian fit with same width as the laboratory ice spectrum (red).}
\label{fig:H2O_ice_YSO}
\end{figure*}

\begin{sidewaystable*}
\tabcolsep=0.1cm
\fontsize{7}{7}\selectfont
\centering
\caption{Ice Column Densities and Abundances}
\label{tab:col density}
\begin{tabular}{l|r|rr|rr|rr|rr|rrr}

\tableline\tableline
\multicolumn{13} {c} {Background Stars} \\

\tableline\tableline

\multicolumn{1}{l}{Source} & \multicolumn{1}{c}{N(H$_2$O)} & \multicolumn{2}{c}{N$_t$(CO) \footnote{Total column density of CO ice.  Where individual components are detected in the CO feature this represents the sum of N$_c$(CO)~+~N$_r$(CO)~+~N$_b$(CO) }} & \multicolumn{2}{c}{N$_c$(CO) \footnote{The central component of the CO ice feature.  This corresponds to the yellow dashed line in Figure \ref{fig:CO_ice}.}} & \multicolumn{2}{c}{N$_r$(CO) \footnote{The long-wavelength (red) component of the CO ice feature.  This corresponds to the red dot-dashed line in Figure \ref{fig:CO_ice}.}}  & \multicolumn{2}{c}{N$_b$(CO) \footnote{The short-wavelength (blue) component of the CO ice feature.  This corresponds to the blue solid line in Figure \ref{fig:CO_ice}.}}  & \multicolumn{3}{c}{N(CH$_3$OH)}  \\ 
\cmidrule(r){3-4}
\cmidrule(r){5-6}
\cmidrule(r){7-8}
\cmidrule(r){9-10}
\cmidrule(r){11-13}
ID & 10$^{18}$ cm$^{-2}$ & 10$^{17}$ cm$^{-2}$ & \%\hto & 10$^{17}$ cm$^{-2}$ & \%\hto & 10$^{17}$ cm$^{-2}$   & \%\hto & 10$^{17}$ cm$^{-2}$   & \%\hto & 10$^{17}$ cm$^{-2}$   & \%\hto & \% CO$_r$  \\

 \tableline
 
B59-B1 & 2.45 (0.46) & $<$14.5 & $<$72.7 &    & &    & … & …    & … &    &  & …     \\
L483-B1 & 0.17 (0.05) & $<$3.3 & $<$275 &     &  & …    & … & …    & … &    &  & …     \\
L483-B2 & 5.34 (0.87) & 12.80 (0.56) & 24.0 (4.0) & 6.14 (0.30) & 11.5 (1.9) & 5.51 (0.37) & 10.3 (1.8) & 1.15 (0.31) & 2.2 (0.7) & 3.01 (0.26) & 5.6 (1.0) & 54.6 (5.9)  \\
L483-B3 & 0.16 (0.03) & 0.62 (0.06) & 38.6 (8.3) &  &  & …    & … & …    & … & $<$1.3    & $<$98.2 & …    \\
L483-B4 & 0.19 (0.04) & $<$0.9 & $<$56.7 &   &  & …    & … & …    & … &    &  & … \\
B335-B1 & 0.78 (0.12) & 4.27 (0.17) & 54.7 (8.7) & 2.82 (0.11) & 36.1 (5.7) & 1.45 (0.13) & 18.6 (3.3) & …    & … & $<$0.7    & $<$11.3 & $<$56.5    \\
L694-B1 & $<$0.17 & $<$1.1 & &     &  & …    & … & …    & … &     &  & …    \\
L694-B2 & 2.10 (0.24) & 11.13 (0.51) & 53.0 (6.5) & 6.57 (0.28) & 31.3 (3.8) & 4.12 (0.33) & 19.6 (2.7) & 0.44 (0.27) & 2.1 (1.3) & 2.98 (0.12) & 14.2 (1.7) & 72.4 (6.5)   \\
L694-B3 & 0.46 (0.07) & 0.52 (0.26) & 11.3 (5.8) &  &  & …    & … & …    & … &    &  & …    \\
\tableline\tableline                              
\multicolumn{13} {c} {Young Stellar Objects} \\
\tableline\tableline

\multicolumn{1}{l}{Source} & \multicolumn{1}{c}{N(H$_2$O)} & \multicolumn{2}{c}{N$_t$(CO) } & \multicolumn{2}{c}{N$_c$(CO)} & \multicolumn{2}{c}{N$_r$(CO)}  & \multicolumn{2}{c}{N$_b$(CO)}  & \multicolumn{3}{c}{N(CH$_3$OH)} \\ 
\cmidrule(r){3-4}
\cmidrule(r){5-6}
\cmidrule(r){7-8}
\cmidrule(r){9-11}
\cmidrule(r){12-13}
ID & 10$^{18}$ cm$^{-2}$ & 10$^{17}$ cm$^{-2}$ & \%\hto & 10$^{17}$ cm$^{-2}$ & \%\hto & 10$^{17}$ cm$^{-2}$   & \%\hto & 10$^{17}$ cm$^{-2}$   & \%\hto & 10$^{17}$ cm$^{-2}$   & \%\hto & \% CO$_r$ \\

\tableline

B59-Y1 & 2.86 (0.36) & 8.50 (0.35) & 29.7 (3.9) & 5.64 (0.19) & 19.7 (2.6) & 2.86 (0.30) & 10.0 (1.6) & …    & … &   $<$2.9  & $<$11.5 &  $<$112.0   \\
B59-Y2 & 0.52 (0.12) & $<$0.9  & $<$21.3 &   &  & …    & … & …    & … & $<$1.3    & $<$31.9 & …     \\
B59-Y3 & 1.76 (0.25) & 2.06 (0.69) & 11.7 (4.3) &  &  & …    & … & …    & … &     &  & …   \\
B59-Y4 & 1.57 (0.23) & 1.55 (0.17) & 9.9 (1.8) &  &  & …    & … & …    & … & $<$2.6    & $<$19.1 &    \\
B59-Y5 & 1.35 (0.21) & 5.05 (0.29) & 37.4 (6.2) & 2.45 (0.14) & 18.1 (3.0) & 2.26 (0.212) & 16.7 (3.1) & 0.35 (0.13) & 2.6 (1.0) & $<$2.1    & $<$18.7 & $<$104.7    \\
\tableline\tableline

\end{tabular}
\end{sidewaystable*}

\subsubsection{CO Ice}\label{sec:COice}
In \citet{Pontoppidan2003} the CO ice absorption feature is broken into three components: the central feature at 4.67 $\mu$m (2139.9 cm$^{-1}$), the blue wing at 4.664 $\mu$m (2143.7 cm$^{-1}$), and the red wing at 4.681 $\mu$m (2136.5 cm$^{-1}$).  The central feature represents pure apolar CO.  The blue wing is most likely formed  when CO mixes with other apolar ices such as CO$_2$ \citep{Elsila1997} while the red wing develops from CO mixing with polar (high-dipole moment) ices \citep{Sandford1988}.  The FWHM of the red component is broad (0.023 $\mu$m or 10.6 cm$^{-1}$) and the central component and blue wing are more narrow (FWHM$_{\text{central}}=$0.0076 $\mu$m or 3.5 cm$^{-1}$ and FWHM$_{\text{blue}}=$0.0065 $\mu$m or 3.0 cm$^{-1}$) \citep{Pontoppidan2003}.   The central and blue features are narrow enough that we must account for broadening due to the spectral resolution of the SpeX instrument.  With a resolution of R$\sim$870 the FWHM of the central CO component and blue component become are 4.28 cm$^{-1}$ and 3.88 cm$^{-1}$ respectively.  

To fit the CO ice features, first a baseline is determined around the feature using a second order polynomial from 4.5 to 4.75 $\mu$m excluding the region between 4.66 and 4.7 $\mu$m.  Then we use the parameters from \citet{Pontoppidan2003} for the central wavelength and FWHM (with the applied corrections for the SpeX instrument) to model the CO ice feature with Gaussian fitting using the IDL routine \texttt{MPFITEXPR}.  The uncertainty incorporates the point-to-point scatter and the IDL routine then returns a 1$\sigma$ error associated with the parameter describing the depth of the absorption feature. The central, red, and blue components are fit simultaneously to determine the individual maximum optical depths ($\tau_{\max}$).  Then we determine the area of the Gaussian as $2.5\times\tau_{\max}\times(FWHM/2.35)$ where the factor of 2.35 comes from the relationship between the FWHM and the Gaussian width. Dividing by the band strength 1.1 $\times$ 10$^{-17}$cm molec$^{-1}$ \citep{Gerakines1995}, we determine the ice column density. 

For seven stars, apolar CO ice is detected above a 3$\sigma$ confidence level and two additional targets have a CO detection above 2$\sigma$ (B59-Y3 and L694-B3) with upper limits calculated for the remaining five targets (See Section \ref{sec:Upper Limts}).  Out of the seven stars we are able to measure the red component for five targets and the blue component for three targets at a 3$\sigma$ level.  These components are further discussed in Section \ref{sec:CO Ice Discussion}.  The optical depth spectrum is displayed in Figure \ref{fig:CO_ice}, normalized to have a baseline of zero, and the three spectral components are shown.

\begin{figure*}[!htb]
\centering
\includegraphics[scale=0.76]{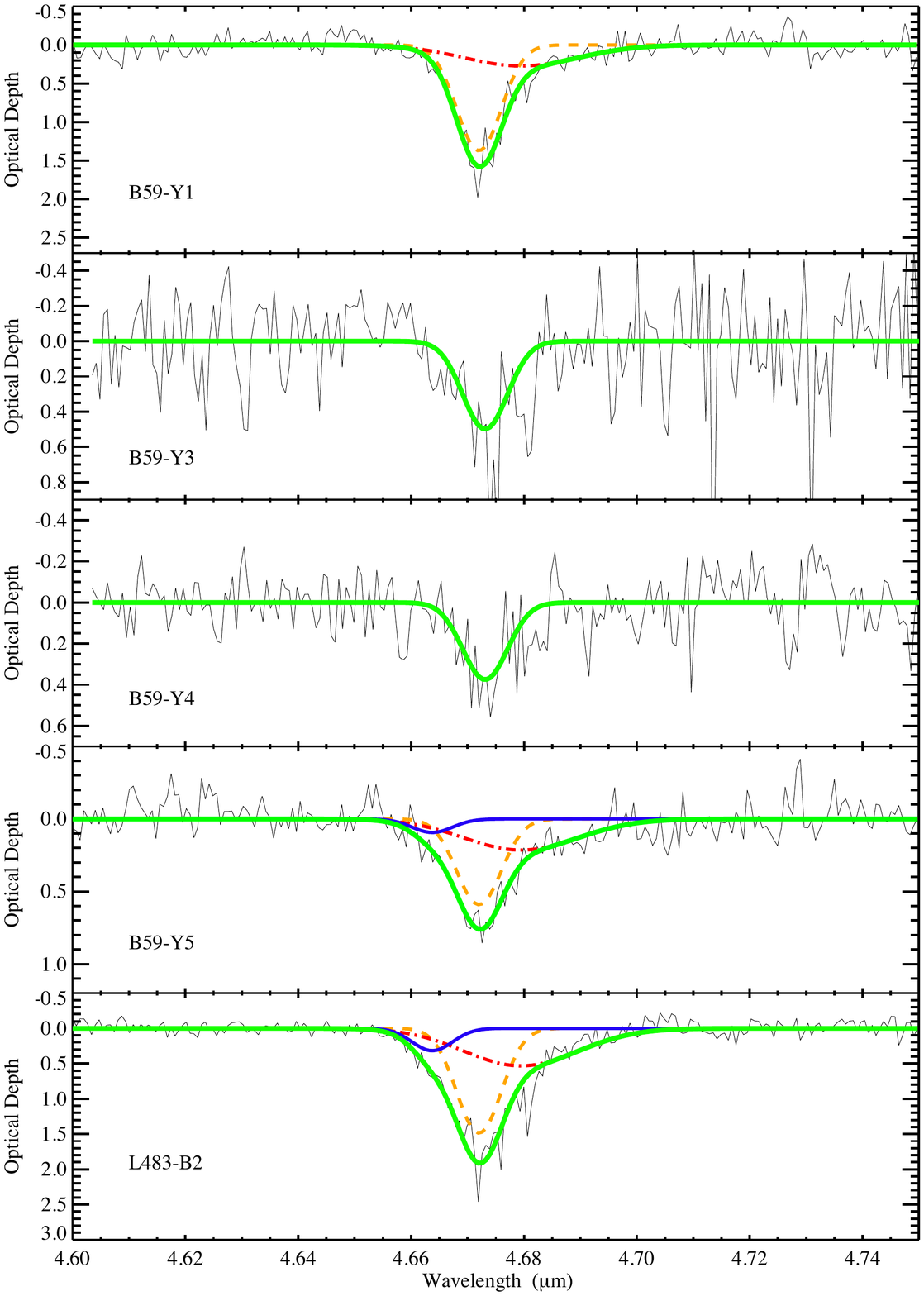}
\caption{CO ice absorption features in the M-band spectrum with the continuum normalized to a zero optical depth (black).  The thick solid green line represents the complete model of the spectral features while the individual components are shown as a blue solid line, an orange dashed line, and a red dot-dashed line (representing the blue, central, and red components respectively).  Only stars that have the CO detections above a 2$\sigma$ level are displayed. }
\label{fig:CO_ice}
\end{figure*}

\begin{figure*}[!htb]
\ContinuedFloat\centering
\includegraphics[scale=0.78]{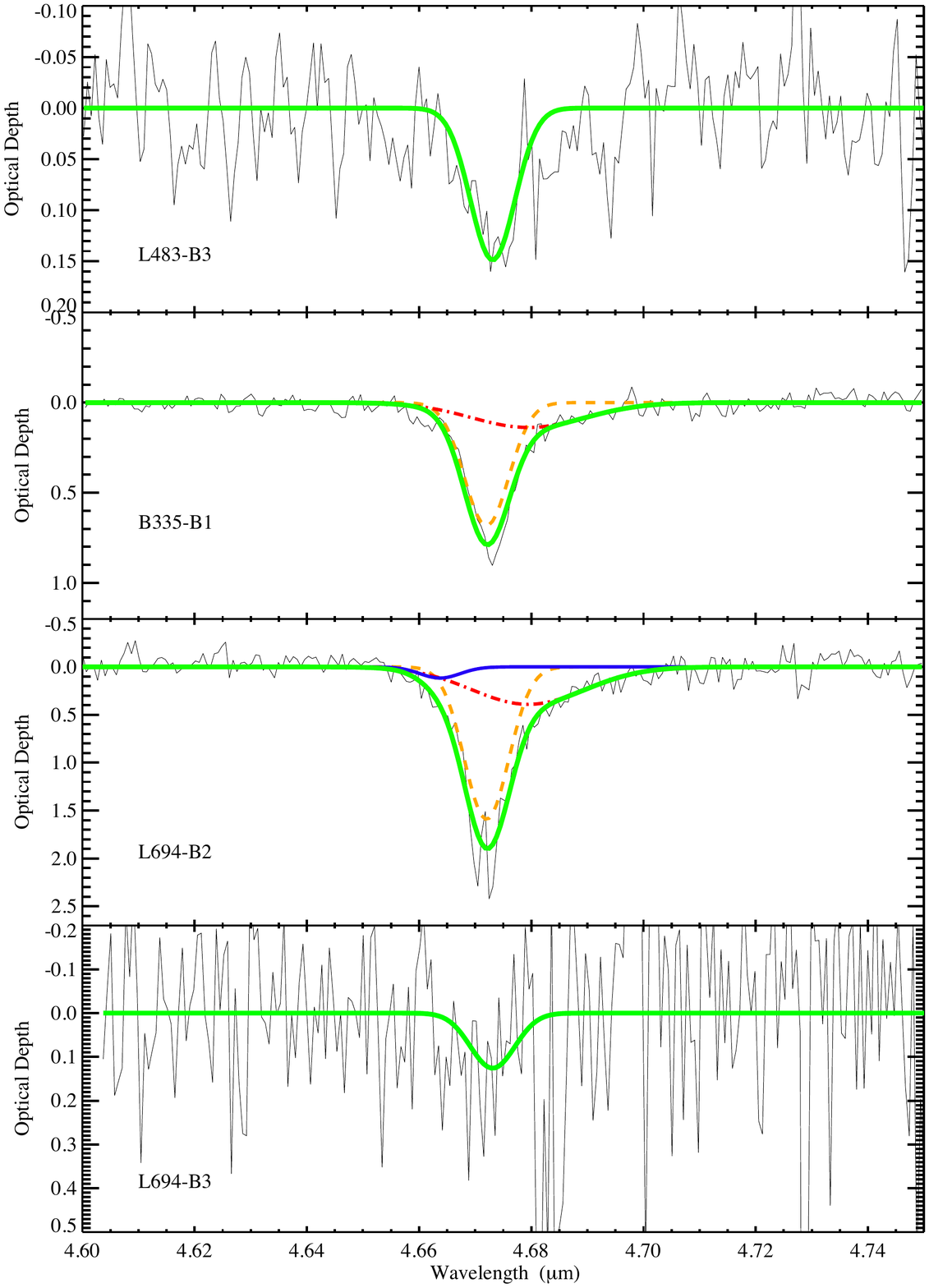}
\caption{(Continued)}
\end{figure*}

\subsubsection{\choh Ice}
\begin{figure*}[h!tb]
\centering
\includegraphics[trim={0 7cm 0 8cm},clip,scale=0.74]{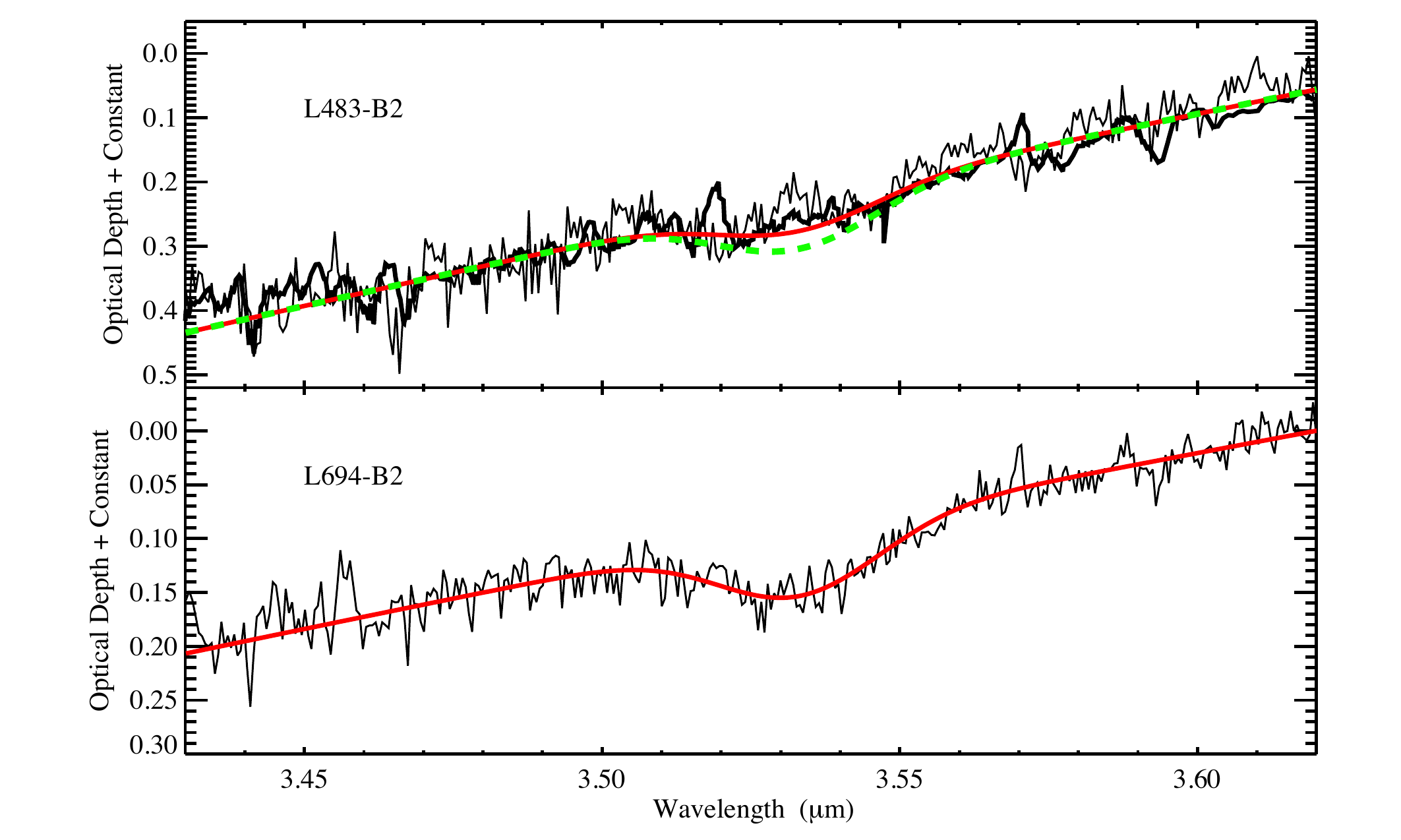}
\caption{\choh ice absorption features in the 3.43-3.62 $\mu$m spectral range with an additional constant in order to set the 3.62 $\mu$m optical depth to zero.  The slope comes from being on the red wing of the \hto absorption feature calculated in this work. Photospheric lines have been subtracted from the spectra using the model templates listed in Table \ref{tab:model_fits}. The thick solid red line represents the  Gaussian model of the absorption feature.  In the top panel the thicker black line represents the spectrum from \citet{Boogert2011} where the green dashed line shows their fit.  The thinner black line spectra is from this work.  These two measurements agree within $\sim$1$\sigma$ according to the errors listed in \citet{Boogert2011}.}
\label{fig:CH3OH_ice}
\end{figure*}

We analyze the \choh ice in the 3.53 $\mu$m C-H stretching mode.  A first order polynomial is fit to the optical depth values from 3.46 to 3.62 $\mu$m excluding the range between 3.5 to 3.56 (to prevent an offset caused by a potential \choh ice feature) and where there are prominent features from the stellar photosphere. Subtracting the polynomial fit from the continuum the star L694-B2 shows a strong visible absorption feature that is fit with a Gaussian profile.  The peak is centered at 2830.3 cm$^{-1}$ (3.533 $\mu$m) with a FWHM of 25.5 cm$^{-1}$ (Figure \ref{fig:CH3OH_ice}).

It is known that the \choh band overlaps with the 3.47 $\mu$m absorption feature, attributed to hydrocarbons \citep{Allamandola1992, Brooke1996} or NH$_3$ hydrates \citep{Dartois2001}.  Therefore, a high quality spectrum is needed to detect the distinct 3.53 $\mu$m absorption peak of the \choh ice feature.  Without a high confidence level in this detection, then the strong contamination with the 3.47 $\mu$m absorption could make any detection of \choh ice very uncertain.  The star L694-B2 has over a 25$\sigma$ confidence thus we use the parameters for the central peak wavelength and FWHM to determine if there are other stars with CH$_3$OH ice detected at lower confidence levels.  Using the same methods as the CO ice, the column density is calculated using a band strength of 5.6~$\times$ 10$^{-18}$cm molec$^{-1}$ \citep{Kerkhof1999}.  

 One additional target (L483-B2) displays \choh ice at above an 11$\sigma$ level.  This star also showed a \choh ice feature in \citet{Boogert2011} (N(CH$_3$OH)=3.63$\pm$0.65). In \citet{Boogert2011} the spectrum was taken using Keck/NIRSPEC from 2.38-4.14 $\mu$m. We took their stellar spectrum and applied the same baseline fitting routine to compare the spectrum with our data.  Figure \ref{fig:CH3OH_ice} shows both spectra on the optical depth scale with the Gaussian fit to the \choh ice feature from both our work and from \citet{Boogert2011}.  The fit is more shallow for our data and the ice abundance is underestimated in our measurements by about a factor of 1$\sigma$.  The spectra match very closely signaling that the discrepancy is not due to short time-scale variability on the ice feature but it is most likely due to a variation in the baseline chosen to measure the feature.

 We determine the fraction of N(CH$_3$OH)/N(H$_2$O) and find this ratio for L483-B2 (5.3 $\pm$ 1.0\%) to be comparable to upper limits for Taurus ($<$3\%) and Lupus ($<$3-8\%). This is lower than the ratio in \citet{Boogert2011} for this target (8.44 $\pm$1.79\%), but is still within 3$\sigma$. For L694-B2 the ratio is 13.3 $\pm$ 1.6\% which is currently the highest fraction detected along the line of sight toward a background star. Some stars in the sample from \citet{Boogert2011} do display ratios above 10\% for isolated dense cores and are within error of this higher ratio. The remaining targets in our sample with high S/N in the continuum emission have calculated upper limits as described in Section \ref{sec:Upper Limts}.

\subsubsection{OCN$^-$ Ice}\label{sec:OCN}

First detected by \citet{Soifer1979}, the 4.62 (2165 cm$^{-1}$) feature commonly known as the "XCN" band was detected toward the massive protostar W33A.  Later this feature was identified by laboratory studies to be the ionic species, OCN$^-$ \citep[\& references therein]{Demyk1998}.  A survey by \citet{vanBroekhuizen2005} shows that the OCN$^-$ abundances toward a sample of 34 deeply embedded low-mass YSOs can vary by an order of magnitude and cannot easily be explained by energetic processes such as UV photons or cosmic rays. It is suggested that grain surface chemistry is responsible for the variations, and \citet{RAUNIER2003} demonstrates through laboratory measurements the possibility of forming OCN$^-$ at cold temperatures (T=10 K).  


Currently no detections of OCN$^-$ have been observed in quiescent clouds or toward background targets \citep{Whittet2001}. This would suggest that it may be necessary to have a YSO environment for OCN$^-$ to form, but we are also limited by the small number of stars that have been observed in this wavelength regime.  We searched for this feature in our sample, and none of our targets (both background stars and YSOs) display the OCN$^-$ feature at 4.62 $\mu$m.  We determine 3$\sigma$ upper limits using the method outlined in Section \ref{sec:Upper Limts} and with the same baseline that was derived for the CO feature.  The FWHM used is 26 cm$^{-1}$ with a band strength of 1.3~$\times$~10$^{-16}$~cm~molec$^{-1}$ \citep{vanBroekhuizen2004,vanBroekhuizen2005}. The upper limits we find ($\sim$2-5$\times$10$^{16}$ molec cm$^{-2}$) are within the range of abundances found in the low mass YSOs in \citet{vanBroekhuizen2005}.  They find abundances from $\sim$0.2-4.9$\times$10$^{16}$ molec cm$^{-2}$ and only one star in our sample has a higher upper limit.  Being limited by sensitivity and a small sample we cannot rule out the possibility of OCN$^-$ formation outside of YSO environments. 

\subsubsection{Upper Limits on Ices}\label{sec:Upper Limts}

In cases where an absorption feature has a significance of less than a 2$\sigma$, the ice column densities are reported as 3$\sigma$ upper limits (Table \ref{tab:col density}).  Because the features are resolved, we first smooth the data using a boxcar method with a width corresponding to half of the FWHM for the feature of interest.  Then the 3$\sigma$ upper limit column density goes as

\begin{equation}
    N < 3\times \sigma \times \text{FWHM} \times A_{\text{bulk}}^{-1}.
\end{equation}

\noindent where $\sigma$ is the standard deviation of the smoothed data and $A_{\text{bulk}}$ is the band strength of the ice species.  The adopted values for $A_{\text{bulk}}$ and FWHM are reported for each ice in the above sections.  The wavelength regimes used to calculate the standard deviation for each molecule are as follows: 2.90--3.20 $\mu$m (H$_2$O), 3.43--3.65 $\mu$m (CH$_3$OH), 4.57--4.66 and 4.70--4.74 $\mu$m (OCN$^-$), and 4.60--4.75 $\mu$m (CO).

\subsection{Correlation Plots}

The correlation of A$_K$ with the $\tau_{\text{3.0}}$ feature has been studied for different environments and cores \citep[e.g.][]{Whittet2001, Chiar1995,Boogert2011, Boogert2013, Murakawa2000}.  Additionally the correlation between A$_K$ with the $\tau_{\text{4.67}}$ feature is presented in \citet{Chiar1995} for the Taurus region. However, the Taurus lines of sight contain little polar-CO, in contrast to several of our sightlines.  Therefore, it is best to compare the relationship of the column density of the total CO (polar and apolar) with A$_K$ rather than $\tau_{\text{4.67}}$, in order to capture all of the CO detected.  We present the correlation of A$_K$ with the column densities for H$_2$O, CO, and CH$_3$OH toward background stars (Figure \ref{fig:Ak_tau}). YSOs are excluded since they are not sampling the full core but only the foreground column and may have a surrounding envelope that could impact the extinction measurement.  


\subsubsection{N(H$_2$O) versus A$_K$}

It has been shown that there is a strong correlation between the peak optical depth (and thus column density) of the H$_2$O ice feature and A$_K$.  Lines of sight toward eight background stars from our study are used to derive a least-squares linear fit with the relation:

\begin{equation}
    N(H_2O)=[(-4.99 \pm 1.86)+(9.15 \pm 0.82)\times A_K]\times10^{17}.
\end{equation}

\begin{figure*}[h!t]
\centering
\includegraphics[trim={0 4cm 0 4cm},clip,width=1\textwidth]{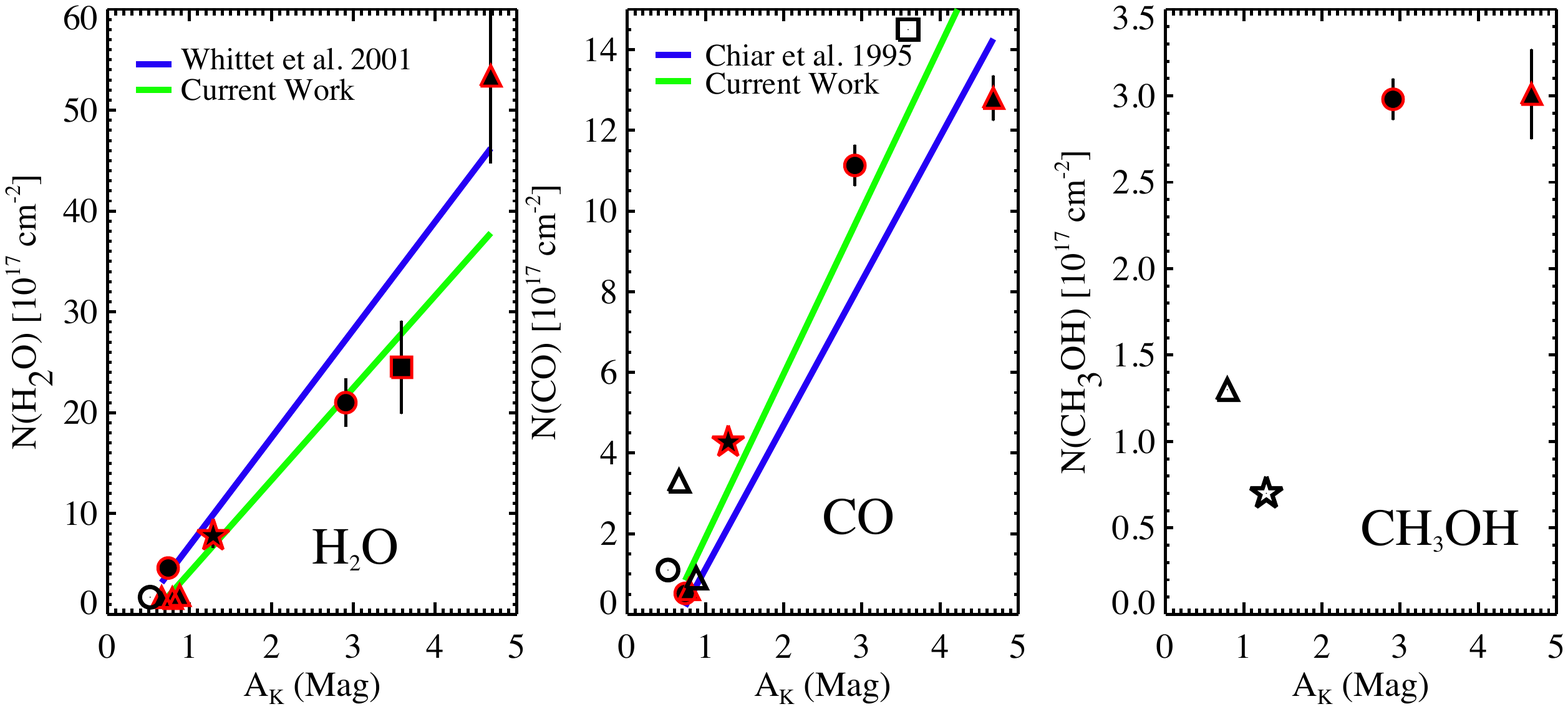}
\caption{The correlation of A$_K$ with the column density from left to right for the H$_2$O feature, the CO feature (N$_t$(CO)) and the CH$_3$OH feature.  Filled data points represent background targets (no YSOs) with a detection above 3$\sigma$ and open points are 3$\sigma$ upper limits.  The symbols represent different clouds: B59 - squares, L483 - triangles, B335 - stars, and L694-2 - circles.  The 1$\sigma$ error bars are shown for the column density, and errors in A$_K$ are smaller than the symbols.  In the left panel the blue line represents the fit found in \citet{Whittet2001} for the Taurus cloud and the green line represents the least-squares fit using the data presented in this work.  In the  center panel the blue line shows the fit also for Taurus stars from \citet{Chiar1995}, and the green line is the least-squares fit to the data presented here.  In the third panel there is no fit to the data since there are only two detections of CH$_3$OH and two upper limits.}
\label{fig:Ak_tau}
\end{figure*}

\noindent To convert this relationship to an optical depth ($\tau_{3.0}$), a factor of 1.7$\times$10$^{18}$ can be divided out. Where $N(H_2O)=0$ we find the cutoff value of A$_K$=0.55$\pm$0.21 or A$_V$=4.87$\pm$1.87 (with A$_V$/A$_K$=8.93, \citet{Rieke1985}). In Figure \ref{fig:Ak_tau} we compare our fit to that of \citet{Whittet2001} for the Taurus cloud, which is accurate and based on a homogeneous sample.  In the following section for the CO ice threshold calculation we also compare to data from Taurus making this a consistent comparison. YSOs are not included in this relationship using the extinction values in Table 1 because the YSOs do not sample the full core as the background stars do.


\subsubsection{N(CO) versus A$_K$}\label{sec:tau CO}

Similarly to the \hto feature, we calculate the correlation of A$_K$ with the column density of the total CO (N$_t$(CO) from Table \ref{tab:col density}) and derive the following relation:

\begin{equation}\label{eqn:CO}
    N(CO)=[(-2.16 \pm 0.36)+(4.06 \pm 0.15)\times A_K]\times10^{17}.
\end{equation}

\noindent At the cutoff value where $N(CO)=0$, we find  A$_K$=0.53$\pm$0.09 or A$_V$=4.75$\pm$0.81.  This is consistent to the relationship found in \citet{Chiar1995} for Taurus, where they find a cutoff of A$_V$=6.0$\pm$4.1 for CO, however this work reports much smaller errors.  Panel 2 of Figure \ref{fig:Ak_tau} displays both fits.



\subsubsection{N(CH$_3$OH) versus A$_K$}

In our sample we detect only two stars with the 3.53 $\mu$m \choh ice feature, and thus we do not fit a correlation (Panel 3 of Figure \ref{fig:Ak_tau}).  Both stars have almost the same abundance of \choh ice despite the differences in extinction, but more sight lines would allow us to determine any trends.  Two stars with upper limits are also shown, and they are much lower than the \choh detections, and they are also at much lower levels of extinction.

\subsection{CO Ice Freeze-out}\label{sec:CO Ice Freeze Out}

Along the lines of sight where we detect CO ice, we can measure the fraction of CO that has frozen out and the CO gas reservoir that remains. The relation N$_H$ = A$_V\times 1.8\times10^{21}$ cm$^{-2}$ can be used where N$_H$ is the total hydrogen column (H and H$_2$)\citep{Predehl1995} and A$_V$/A$_K$=8.93\citep{Rieke1985}.  The relationship between the molecular hydrogen and CO column density in dense clouds was measured by \citet{Lacy1994} as 4000$\pm$3000, and since then other work has shown similar values but with large uncertainties \citep[summarized by][]{Lacy2017}.  We adopt the relation N$_{H_2}$/N$_{CO}$ = 5000 and thus N$_{CO}$/N$_H$=1$\times10^{-4}$ (with N$_H$=2$\times$N$_{H_2}$) for the measurement of the total amount of N$_{CO}$ (gas and ice) within the column.  In Table \ref{tab:frozen_ice} the fraction of CO$_{\text{ice}}$:CO$_{\text{total}}$ and CH$_3$OH$_{\text{ice}}$:CO$_{\text{total}}$ is reported.  The CO$_{\text{ice}}$ ratios for all of the targets are low, less than 15\%. 

\begin{table*}[h!t]
\tabcolsep=0.1cm
\tabletypesize{\small}
\centering
\caption{Fraction of Frozen Out Ice}
\label{tab:frozen_ice}
\begin{tabular}{lcc}

\tableline\tableline

Source ID & CO$_{ice}$:CO$_{tot}$ & CH$_3$OH$_{ice}$:CO$_{tot}$ \\
\tableline
B59-Y1 & 0.061 & \\
B59-Y3 & 0.034 & \\
B59-Y4 & 0.085 & \\
B59-Y5 & 0.069 & \\
L483-B2 & 0.080 & 0.033 \\
L483-B3 & 0.043 & \\
B335-B1 & 0.136 & $<$ 0.018 \\
L694-B2 & 0.140 & 0.064 \\
L694-B3 & 0.044 & \\

\tableline

\end{tabular}
\end{table*} 

\section{Discussion: The Relationship of CO and CH$_3$OH Ice}\label{sec:Discussion}\label{sec:CO Ice Discussion}

For the first time, we have simultaneously measured the CO and \choh ice features for background targets in dense cores to observationally constrain the CO freeze-out process.  From theoretical work, CO undergoes almost a complete freeze out at low temperatures and high densities ($n$ $\geq$ 10$^5$ cm$^{-3}$) leading to CO hydrogenation producing H$_2$CO and CH$_3$OH \citep{Cuppen2009}. The quiescent regions in the Taurus cloud do not show this freeze-out, but dense and prestellar cores indicate that this phase has begun before stars form (\citet{Boogert2015} and references therein).

Along our lines of sight only a small amount of CO is frozen out ($\leq$15\%, Table \ref{tab:frozen_ice}).  The \choh ice is likely mixed with CO based on the detection of the polar CO signature along lines of sight where \choh is detected.  This layer is CH$_3$OH-rich (CH$_3$OH/CO$_r>50$\%; Table \ref{tab:col density}).  For lines of sight with CO$_r$ detections but \choh non-detections, the upper limits for \choh are still consistent with a large CH$_3$OH:CO$_r$ ratio.  The \choh ice constitutes only a small fraction of the core along the line of sight ($\lesssim$8\% by total CO mass).  This means that the columns are mostly dominated by the lower extinction regions of the core where CO still remains in the gas phase.  Pure CO ice likely traces high density regions, but presumably only in the densest regions can \choh ice form. Noteably, the target with the highest abundance of frozen CO (L694-B2) also has the highest \choh ice abundance.  Without more stars this cannot be identified as a correlation due to potential environmental variations in other cores.


\subsection{Comparison to Previous Ice Measurements}

\begin{figure}[h!t]
\centering
\includegraphics[width=1\textwidth]{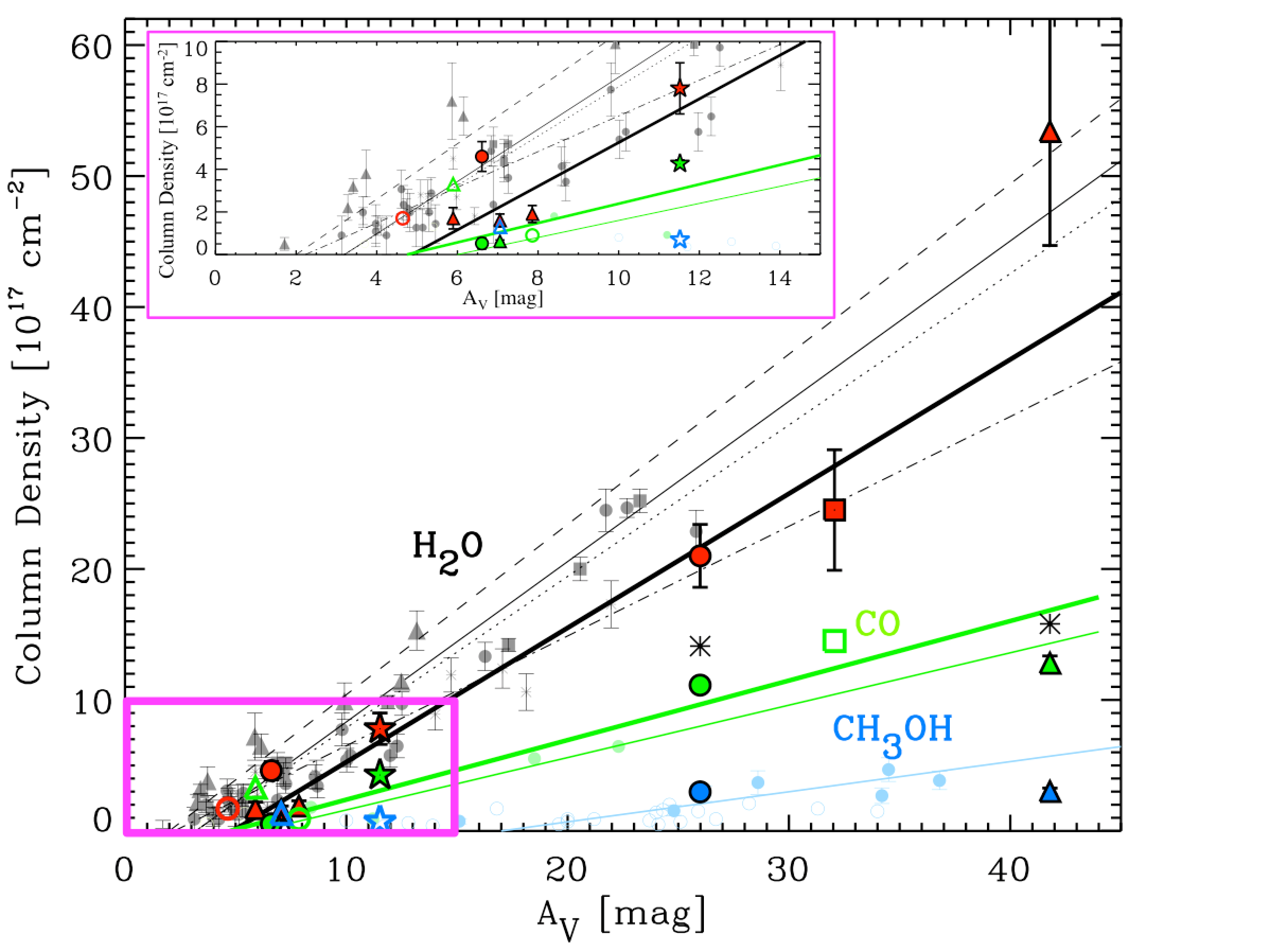}
\caption{The relationship of the \hto (grey and red), CO (green) and \choh (blue) ice column densities with the extinction A$_V$ for background stars tracing nearby clouds and cores.  Data presented in this work is overlaid on data from previous studies.  Open symbols are upper limits and the lines represent linear fits to the ice detections.  Our data follow the same symbols as Figure \ref{fig:Ak_tau}.    Dark grey points represent previous data from Taurus (circles and solid thin line) and Lupus IV (asterisks and dash-dot line) clouds, as well as the L183 (triangles and dashed line) and IC 5146 (squares and dotted line) cores.  We display the \hto column densities from this work as red symbols and our fit is the thick black solid line.  CO column densities from Taurus are represented by small green circles and our data is shown with larger green symbols with black outlines and a thick green line representing the best fit. For CO the total column density is shown as the sum of the individual components (N$_t$(CO) from Table \ref{tab:col density}). For \choh previous data comes from a variety of dense cores and are represented as small light blue circles with a derived fit.  Our data are represented by larger blue symbols with black outlines.  Large black asterisks are shown for the two stars that have \choh detections as a sum of the CO and \choh ice column densities to represent the amount of CO that would exist if it were not converted to CH$_3$OH (see Section \ref{sec:ice conversion}). The previous data shown is taken from \citet{Boogert2015} and references therein.  The inset figure shows a close-up of low A$_V$ (0-15 mag) region highlighted in magenta where it is difficult to distinguish between different data points.}  
\label{fig:boogert plot}
\end{figure}

In \citet{Boogert2015}, a compilation of background targets to different molecular cores show the column densities of H$_2$O, CO, and \choh as a function of extinction (A$_V$).  In Figure \ref{fig:boogert plot} we overlay our background stars for comparison on the relationship and the ice formation thresholds.

The \hto threshold found in this work (A$_V$=4.87$\pm$1.87) is within 1$\sigma$ of that found for several other cores (e.g. \citet{Chiar1995} A$_V$=3.1$\pm$0.6, \citet{Whittet2001} A$_V$=3.2$\pm$0.1, \citet{Boogert2011} A$_V$=2.9$\pm$2.6, \citet{Boogert2013} A$_V$=2.1$\pm$0.6).  Other regions have a broad range of extinction thresholds such as the Pipe nebula (excluding B59) at A$_V$=5.2$\pm$6.1 \citep{Goto2018} and Ophiuchus at A$_V$=10-15 \citet{Tanaka1990}.  In \citet{Boogert2011}, where a variety of cores with nearly 30 background stars are used to derive the \hto ice threshold a large uncertainty is also found.  This is explained by the clouds experiencing different environmental effects including external radiations fields, dust temperature, and density, particularly because our sample has both collapsing and star forming cores. 

\subsection{CO to CH$_3$OH Ice Conversion}\label{sec:ice conversion}

This is the first time CO ice has been measured for background stars with extinctions above A$_K \sim 2.6$. It may be that a linear fit as shown in Figure \ref{fig:Ak_tau} is no longer applicable as some CO is converted into \choh ice. The conversion of CO into \choh is evident for the star tracing L483 (L483-B2) where at the high extinction, A$_K$=4.68, the column density is lower than the derived relationship in Equation \ref{eqn:CO}.  In Figure \ref{fig:boogert plot}, asterisks represent the sum of CO and CH$_3$OH column densities, and show that L483-B2 would then match the best fit for CO.  However, for L694-B2, the addition of CH$_3$OH ice removes the source further from the value expected from the linear relation.

The conversion of CO into \choh in quiescent clouds is significant and could be sufficient to explain the \choh that is detected in Class 1 low mass YSO envelopes.  In Table \ref{tab:yso comparison} we compile a list of low and high mass YSOs and find the ratios of CH$_3$OH:CO$_{\text{r}}$ for the Class 1 YSOs to be comparable to the ratios we find for the stars tracing both L483 and L694-2 where the ratio for L694-2 is remarkably even higher than the low mass Class 1 YSOs (Table \ref{tab:col density}).  In the envelopes of some high mass YSOs more \choh ice is produced but the wide variation most likely reflects environmental variations in the local conditions such as the densities and dust temperatures.  The Class 0 star in Table \ref{tab:yso comparison} shows a much higher abundance.  More observations are needed of these very young objects to determine any conclusive trends with evolutionary stage.

\begin{table*}[h!t]
\tabcolsep=0.1cm
\tabletypesize{\small}
\centering
\caption{Ice Column Densities and Ratios for Literature YSOs}
\label{tab:yso comparison}

\begin{tabular}{lcccccc}
\tableline\tableline

Star & Mass Range & Classification & N$_t$(CO)\footnote{\label{note1}*\citet{Boogert2002}, \dag \citet{Boogert2004}, \ddag \citet{Pontoppidan2003}, \S \citet{Pontoppidan2008}, ** \citet{Boogert2008}, \dag\dag \citet{Pontoppidan2004}, \ddag\ddag \citet{Dartois1999}}\footnote{The total CO column density combining polar and apolar components} & CO$_{\text{c+b}}$:CO$_{\text{r}}$\footref{note1}\footnote{\label{note2}\footnote{CO$_{\text{c+b}}$ represents the apolar ice combining the central and blue components} CO$_{\text{r}}$ represents CO mixed with polar ice} & N(CH$_3$OH)\footref{note1}	 & CH$_3$OH:CO$_{\text{r}}$ \footref{note2}\\
 & & & 10$^{17}$ cm$^{-2}$ & & 10$^{17}$ cm$^{-2}$ & \\

\tableline	
L1489IRS	&	Low	&	Class I	&	6.50 (0.65)	*	&	0.86 (0.05)	*	&	2.10 (0.67)	**	&	0.60 (0.20)	\\
HH46	&	Low	&	Class I	&	10.3 (0.80)	\dag	&	0.34 (0.05)	\dag	&	4.32 (0.31)	**	&	0.56 (0.04)	\\
CrA 5	&	Low	&	Class I	&	16.1 (0.61)	\ddag	&	2.98 (0.20)	\ddag	&	2.00 (0.35)	**	&	0.50 (0.09)	\\
SVS4-5/EC90	&	Low	&	Class 0	&	16.6 (1.4)	\ddag	&	0.61 (0.12)	\ddag	&	14.2 (2.0)	\dag\dag	&	1.38 (0.23)	\\
GL2136	&	High	&	-	&	2.25 (0.51)	\S	&	0.17 (0.09)	\S	&	3.90 (0.97)	\ddag\ddag	&	2.03 (0.68)	\\
GL989	&	High	&	-	&	3.44 (0.36)	\S	&	0.88 (0.18)	\S	&	0.67 (0.06)	**	&	0.37  (0.05)	\\
W33A	&	High	&	-	&	9.10 (1.00)	\ddag	&	0.26 (0.06)	\ddag	&	18.5 (4.62)	\ddag\ddag	&	2.56 (0.70)	\\
GL7009S	&	Hgih	&	-	&	11.2 (0.40)	\ddag	&	0.31 (0.04)	\ddag	&	35.5 (8.8)	\ddag\ddag	&	4.15 (1.05)	\\
NGC7538IRS9	&	High	&	-	&	17.0 (2.0)	*	&	2.38 (0.20)	*	&	4.85 (1.21)	\ddag\ddag	&	0.96 (0.24)	\\

\tableline\tableline                              
\multicolumn{7} {c} {Stars From This Work} \\
\tableline\tableline
L483-B2 & - & Background & 12.80 (0.56) & 1.32 (0.12) & 3.01 (0.26) & 0.55 (0.06) \\
L694-B2 & - & Background & 11.13 (0.51) & 1.70 (0.17) & 2.98 (0.12) & 0.73 (0.07) \\
\tableline

\end{tabular}
\end{table*}

To understand the types of environments the \choh ice can develop in, we look more closely at where \choh ice is detected in L483 and L694-2. For L483, the environment contains a deeply embedded (A$_V\sim70$) protostar between the Class 0 and Class I stage \citep{Tafalla2000, Fuller1995}.  The line of sight that we trace is at a lower extinction than the embedded YSO (A$_V\sim40$), and its projected distance on the sky is $\sim$14,000-18,000 AU away from the protostar in the northwest direction (similar to the distance to the Oort cloud in our own Solar System).  In Figure \ref{fig:color images} it is clear that the background star is nearly orthogonal to the bipolar outflows and thus the ice growth should not be affected by the enhanced UV field or shocks from the YSO.  Preliminary extinction maps show that the background star does not appear to be sampling a different clump from the protostar as the extinction map shows a nearly spherical extinction structure around the protostar and the background star traces an outer shell (Chu et al. in prep).  Assuming turbulent fragmentation to form individual cores for low mass stars, we can assume YSO envelope size scales up to the Bonnor-Ebert radius (10$^4$ AU for $n_H=10^5$ cm$^{-3}$, $T=$10K, \citet{Offner2010}).  This means the background star is not part of the YSO envelope and the \choh ice is forming without influence from the nearby YSO.

The environment for L694-2 is different in that it is a starless core with a cometary tail-like shape extending to the southeast from the densest part.  The line of sight where we detect \choh ice lies within this extended feature very near the edge of the core where background sources become obscured in the K band about 0.1 pc (20,000 AU) away from the densest region.   A similar star tracing the starless collapsing (or near-collapsing) core L429-C also shows \choh ice outside the densest part of the core \citep{Boogert2011, Stutz2009}.  

While we do not directly observe more complex molecules in the ices, the very high conversion rate of CO to \choh indicates that, most likely, even in the pre-stellar phase there is an abundance of complex organic molecule (COM) formation.  It was experimentally shown that intermediate radicals produced along the CO to \choh hydrogenation route will lead to COMs such as glycolaldehyde, ethylene glycol and, possibly, glyoxal.  This happens through "non-energetic" reactions on grain surfaces and could dominate in dark clouds \citep{Fedoseev2015, Butscher2015, Chuang2016, Fedoseev2017}.  In later stages of the star and planet formation "energetic processing" is possible where \choh is irradiated by energetic particles such as protons, electrons, X-rays, and VUV-photons.  Many laboratory studies have proven that COMs are efficiently formed through recombination of carbon bearing \choh dissociation products \citep[e.g.][]{Allamandola1988, Hudson2000, Oberg2009, Henderson2015, Chuang2017}.  Observationally, the distinction between these two formation mechanisms is not proven, but our observations with abundant \choh ice favor the non-energetic process at low temperatures, in the prestellar phase. It cannot yet be determined if more \choh is formed in the Class 0 and 1 stages by energetic processing, but clearly much of the \choh ice is delivered from the dense cloud to the protostellar envelope without the need for energetic processes.



\subsection{Polar CO Without CH$_3$OH Ice Feature}

Three lines of sight in our sample have polar CO ice detected without an independent \choh ice detection. The column densities of the polar CO are lower by more than 30\% of those discussed in the previous section that trace L483 and L694-2. Two stars trace B59 and are identified as Class 2 YSOs with comparable levels of extinction to those where \choh ice is found (and thus high densities).  The absence of \choh is most likely from low sensitivity as the upper limits on the \choh ice band are not significantly lower than our \choh ice detections.  The variations in the environment could also affect the \choh ice development.  Reported turbulence in B59 perhaps has limited the CO freeze out \citep{DuarteCabral2012}.

The third star with polar CO and no independent \choh ice measurement traces B335.  The extinction is lower than the lines of sight in L483 and L694-2 so at this lower density the ice hydrogenation stage could be just beginning and \choh ice is not overly abundant.  The upper limit is more than 3$\sigma$ lower than our lowest \choh detection so if it were abundant at this level of extinction it should have been detected.


\section{Summary}

We have observed the spectra from 2-5 $\mu$m for 9 background targets and 5 YSOs sampling four small  dense molecular cores.  The H$_2$O, CO, and \choh ices were studied to observationally constrain the onset of complex organic molecule formation.  We find that in the cores studied only a small amount of CO is frozen out ($\leq$ 15\%) and in 2 cores, $\sim$30\% of the CO ice is in the polar environment, most likely due to mixing with \choh, presumably tracing the highest densities.  However, \choh is found in abundance along the sight lines where it is detected and is mostly present in a CH$_3$OH-rich CO ice layer.  

The conversion of CO ice into \choh ice is sufficient to explain the \choh ice that is detected toward low mass Class 1 YSO envelopes. However, the environments where ice is detected can affect the efficiency of the CO to \choh ice conversion where factors such as dust densities and temperatures play a role.  We do not directly observe complex organic molecules (COMs), but the high CO to \choh conversion rate implies that COMs can form even before the development of a star.  This depends upon models where \choh forms under cold conditions without energetic processing.

This work will be greatly expanded upon with the launch of the James Webb Space Telescope (JWST) where fainter stars can be detected in the 3-5 $\mu$m range.  This will provide a much larger sample to study the onset of COMs and map the environments that allow ices to form and the evolution of their growth.

{\it Acknowledgements} -- We would like to extend our appreciation to the anonymous reviewer who greatly improved the clarity of this manuscript. We thank Jason Chu for helpful discussions during data analysis.  This material is based upon work supported by the National Aeronautics and Space Administration (NAS5-02105) and by the Spitzer Space Telescope  (PID 11028).  This publication makes use of data products from the Wide-field Infrared Survey Explorer, which is a joint project of the University of California, Los Angeles, and JPL/Caltech, funded by NASA. This publication also uses data products from the Two Micron All  Sky Survey, which is a joint project of the University of Massachusetts and the Infrared Processing and Analysis Center/Caltech, funded by NASA and the National Science Foundation.  We also thank the Soroptimist International Founder Region Fellowship for Women for their generous contribution supporting this work.  The authors recognize that the summit of Maunakea has always held a very significant cultural role for the indigenous Hawaiian community. We are thankful to have the opportunity to use observations from this mountain.

\bibliographystyle{aasjournal}
\bibliography{references.bib}

\end{document}